\documentclass[aps,prl,twocolumn,amsmath,amssymb,nofootinbib,longbibliography,superscriptaddress]{revtex4}
\usepackage{times}
\usepackage[english]{babel}
\usepackage{latexsym}
\usepackage{graphics}
\usepackage{subfigure}
\usepackage{epsfig}
\usepackage{color}
\usepackage{hyperref}
\usepackage{braket} 
\usepackage[T1]{fontenc}
\usepackage[latin9]{inputenc}
\usepackage{amstext}
\usepackage{amssymb}
\usepackage{graphicx}
\usepackage{esint}
\usepackage{braket}
\usepackage{babel}
\usepackage{amsmath}

\hypersetup{
	colorlinks=true,
	citecolor=blue,
	linkcolor=red,
	urlcolor=black
}


\newcommand{\Er}{E_{\textrm{r}}}

\newcommand{\fc}{f_{\textrm{c}}}

\begin{document}
\newcommand{\Jnature}{Nature (London)}
\newcommand{\Jnatphys}{Nat. Phys.}
\newcommand{\Jnatcomm}{Nat. Comm.}
\newcommand{\Jnatmat}{Nat. Mater.}
\newcommand{\JSciRep}{Sci. Rep.}

\newcommand{\Jscience}{Science}
\newcommand{\Jsciadv}{Science Advances}

\newcommand{\Jpnas}{Proc. Nat.l Acad.  Sci.}

\newcommand{\Jprx}{Phys. Rev. X}
\newcommand{\Jprl}{Phys. Rev. Lett.}
\newcommand{\Jpr}{Phys. Rev.}
\newcommand{\Jpra}{Phys. Rev. A}
\newcommand{\Jprb}{Phys. Rev. B}
\newcommand{\Jprc}{Phys. Rev. C}
\newcommand{\Jprd}{Phys. Rev. D}
\newcommand{\Jpre}{Phys. Rev. E}
\newcommand{\Jrmp}{Rev. Mod. Phys.}

\newcommand{\JplA}{Phys. Lett. A}

\newcommand{\Jepl}{Europhys. Lett.}
\newcommand{\Jnjp}{New J. Phys.}
\newcommand{\Jepjb}{Eur. Phys. J. B}
\newcommand{\Jepjd}{Eur. Phys. J. D}
\newcommand{\Jepjst}{Eur. Phys. J. Special Topics}

\newcommand{\crasphy}{C. R. Phys.}

\newcommand{\Jjosab}{J. Opt. Soc. Am. B}
\newcommand{\JApplPhysLett}{Appl. Phys. Lett.}
\newcommand{\JApplPhysB}{Appl. Phys. B}

\newcommand{\Joptcomm}{Opt. Comm.}
\newcommand{\JApplOpt}{Appl. Opt.}

\newcommand{\JphysA}{J. Phys. A}

\newcommand{\JphyslettA}{J. Phys. Lett. A}
\newcommand{\JphyslettB}{J. Phys. Lett. B}

\newcommand{\JnulcphysA}{Nucl. Phys. A}

\newcommand{\Jbullamphyssoc}{Bull. Am. Phys. Soc.}

\newcommand{\Jcpl}{Chin. Phys. Lett.}

\newcommand{\Jphysjap}{J. Phys. Soc. Japan}

\newcommand{\JJlowT}{J. Low Temp. Phys.}

\newcommand{\Jprocroysoc}{Proc. Roy. Soc. A: Math. Phys. Eng. Sci.}

\newcommand{\Jjetp}{Sov. Phys. JETP}
\newcommand{\Jjetplett}{JETP Lett.}
\newcommand{\JjphysUSSR}{J. Phys. USSR}
\newcommand{\JSovJlowT}{Sov. J. Low Temp. Phys.}

\newcommand{\JZhEkspTeorFiz}{Zh. Eksp. Teor. Fiz.}

\newcommand{\Jjchemphys}{J. Chem. Phys.}
\newcommand{\Jjphyschemsol}{J. Phys. Chem. Sol.}
\newcommand{\Jsolstatecomm}{Solid State Comm.}

\newcommand{\Jijmpb}{Int. J. Mod. Phys. B}

\newcommand{\Jijthphys}{Int. J. Theor. Phys.}

\newcommand{\JphysicaB}{Physica B}
\newcommand{\JphysicaBC}{Physica B+C}

\newcommand{\Jstatmech}{J. Stat. Mech.}
\newcommand{\Jstatphys}{J. Stat. Phys.}

\newcommand{\Jphysrep}{Phys. Rep.}
\newcommand{\JRepProgPhys}{Rep. Prog. Phys.}
\newcommand{\JjphysCM}{J. Phys.: Cond. Matt.}
\newcommand{\JjphysA}{J. Phys. A: Math. Theor.}
\newcommand{\JjphysB}{J. Phys. B: At. Mol. Opt. Phys.}
\newcommand{\JjphysC}{J. Phys. C: Solid State Phys.}
\newcommand{\JjphysF}{J. Phys. F: Metal Phys.}
\newcommand{\Jphystoday}{Phys. Today}

\newcommand{\Jadvphys}{Adv. Phys.}

\newcommand{\Jannphys}{Ann. Phys. (NY)}

\newcommand{\JAnnualRevCondMat}{Annual Rev. Cond. Mat. Phys.}

\newcommand{\Jadvatmoloptphys}{Adv. At. Mol. Opt. Phys.}
\newcommand{\Joptexpr}{Opt. Express}

\newcommand{\JphysB}{J. Phys. B: At. Mol. Opt. Phys.}

\newcommand{\JTheorMathPhys}{Theor. Math. Phys.}
\newcommand{\JMathPhys}{J. Math. Phys.}
\newcommand{\JCommMathPhys}{Comm. Math. Phys.}

\newcommand{\Jprogthphys}{Prog. Theor. Phys.}
\newcommand{\Jprogthphyssup}{Prog. Theor. Phys., Suppl.}

\newcommand{\JjphysFr}{J. Phys. (France)}
\newcommand{\JjphysquatreF}{J. Phys. IV (France)}

\newcommand{\Jcommmathphys}{Comm. Math. Phys.}

\newcommand{\JZphys}{Z. Phys.}
\newcommand{\JZphysB}{Z. Phys. B}

\newcommand{\JRevSciInstrum}{Rev. Sci. Instrum.}

\newcommand{\JFortschrPhys}{Fortschr. Phys.}

\newcommand{\Jcompphyscom}{Comput. Phys. Commun.}
\newcommand{\Jphysconfser}{J. Phys. Conf. Ser.}

\newcommand{\Jphysicascripta}{Phys. Scr.}

\newcommand{\Jphysics}{Physics}

\title{
Two-mode collapse and revival of quantum coherent state in a tilted optical lattice
}

\author{Chi-Kin Lai}
\thanks{These authors contributed equally to this work.}
\affiliation{State Key Laboratory of Photonics and Communications, School of Electronics, Peking University, Beijing 100871, China}
\author{Shengjie Jin}
\thanks{These authors contributed equally to this work.}
\affiliation{International Center for Quantum Materials, School of Physics, Peking University, Beijing 100871, China}
\author{Yuanzhe Hu}
\thanks{These authors contributed equally to this work.}
\affiliation{State Key Laboratory of Photonics and Communications, School of Electronics, Peking University, Beijing 100871, China}
\author{Zhongshu Hu}
\affiliation{International Center for Quantum Materials, School of Physics, Peking University, Beijing 100871, China}
\author{Fansu~Wei}
\affiliation{State Key Laboratory of Photonics and Communications, School of Electronics, Peking University, Beijing 100871, China}
\author{Congwen Li}
\affiliation{State Key Laboratory of Photonics and Communications, School of Electronics, Peking University, Beijing 100871, China}
\author{Tianwei Zhou}
\affiliation{Department of Physics and Astronomy, University of Florence, 50019 Sesto Fiorentino, Italy}
\author{Hepeng Yao}\email{hepeng.yao@pku.edu.cn}
\affiliation{State Key Laboratory of Photonics and Communications, School of Electronics, Peking University, Beijing 100871, China}
\author{Xiaoji Zhou}\email{xjzhou@pku.edu.cn}
\affiliation{State Key Laboratory of Photonics and Communications, School of Electronics, Peking University, Beijing 100871, China}

\date{\today}
	
\begin{abstract}

Collective dynamics is an important out-of-equilibrium feature of quantum coherent states and usually reflects the intrinsic properties of the state. Collapse and revival (CR) dynamics of phase coherence is a well-known example for bosonic coherent states, which is usually induced by applying a quench. Previous studies have shown that the CR frequency is governed solely by interactions, even in the presence of a tilt quench. However, whether such interaction-dominated oscillation is a universal feature remains unknown. In this work, we show that an ensemble of one-dimensional bosons can undergo two-mode CR, with frequencies set by both the interaction and the tilt, particularly when the tilt is weaker than the interaction. The newly discovered tilt mode is enabled by tunneling between lattice sites. When the two modes coexist, the amplitudes of both modes exhibit universal linear scaling for various tilts. These findings clarify the general features of CR dynamics in tilted lattice models and the underlying mechanism, and provide deeper insight into collective dynamics in correlated systems.
\end{abstract}
\maketitle
	
\section*{INTRODUCTION}
Phase coherence is a fundamental concept in physics, playing a crucial role across a wide range of fields, including wave mechanics, optics, and quantum mechanics. It underlies phenomena such as the propagation of acoustic waves~\cite{2020_AcousticWaveNonreciprocity,2026AcousticWavePlasmon,wendt2026electrically}, interference in Young's double-slit experiment~\cite{zhou2021doubleslit,tirole2023double,hong2025fetosecond}, and coherent light-atom interactions in cavity quantum electrodynamics~\cite{PhysRev.140.A1051,PhysRevLett.44.1323,PhysRevLett.58.353,PhysRevLett.76.1800}. In quantum mechanics, thanks to the wave-particle duality, the motion of particles is always described by their wave functions, where the phase factor underpins quantum interference and entanglement~\cite{andrews1997twoBEC,greve2022entanglement,Chakraborti2025grapheneinterferometer,pedalino2026probing,Li2026atomicsensors}. For a many-body quantum system, when the phases of individual particles become coherently "locked" together, they collectively form a macroscopic quantum state. Remarkably, they can be effectively described within the framework of coherent quantum states~\cite{PhysRev.131.2766,klauder1985coherent,RevModPhys.62.867}, including Bose-Einstein condensates (BECs), Schr\"odinger cat states, lasers and superconductors~\cite{PhysRev.108.1175,2013VlastakisCatState}. These coherent states are usually stable and possess long lifetimes. Consequently, they are widely used in quantum optics, quantum information, and precision measurement.

From the dynamical perspective, one important feature of coherent quantum state is the collective dynamics, which usually happens when such systems are prepared in a superposition of number states~\cite{PhysRevLett.77.2158,PhysRevLett.78.2511}.
One typical example is a BEC in optical lattices.
By applying a quench (i.e., a rapid increase) in lattice depth, a collapse and revival (CR) dynamics happens where phase coherence is periodically lost and restored with a frequency set by the two-body interaction~\cite{Greiner2002,PhysRevA.74.053616,PhysRevLett.98.180601,PhysRevLett.98.200405,PhysRevA.99.013602}. 
With improved homogeneity and longer measurement times, one can even detect the effective higher-order interactions, which are typically an order of magnitude smaller than the two-body interaction~\cite{Johnson_2009,Will2010,PhysRevA.83.063609}.

Tilted lattice model is another typical example, where a linear gradient potential (tilt) coexists with the optical lattice. By quenching the tilt, the collective dynamics appears as a combination of CR dynamics and Bloch oscillations (BO)~\cite{PhysRevA.89.023606,PhysRevLett.90.213002,PhysRevLett.112.193003}. The latter correspond to periodic motion in quasi-momentum space under a constant force~\cite{PhysRevLett.76.4508,PhysRevLett.97.060402,PhysRevLett.120.213201,GUO20222291,rabec2025bloch}.
In known theoretical and experimental observations~\cite{PhysRevA.89.023606,PhysRevLett.90.213002,PhysRevLett.112.193003}, the probed CR frequency remains set solely by the interaction energy, regardless of the tilt. In addition to BO and CR, tilted lattice models provide a versatile platform for exploring a wide range of phenomena~\cite{Trotzky2008,PhysRevLett.124.043204,Alexander2022,PhysRevA.70.015604,Simon2011,PhysRevLett.111.053003,Zhu2024}, such as the realization of fractional quantum Hall states~\cite{Leonard2023} and the simulation of anyonic models~\cite{Dhar2025,2np8-mp39}. Therefore, given these rich applications and broad interest, it is important to understand the underlying CR mechanism in tilted lattice systems. 
In particular, for both of the two examples above, 
the inter-site tunneling is suppressed and only the on-site interaction mode exists. This naturally raises a question: can the CR dynamics originate from modes controlled by inter-site process? Or is such dynamics solely dominated by on-site interactions a universal feature? From a broader perspective, it is crucial to understand the general behavior of such collective dynamics.

In this work, we provide the first observation of two-mode CR dynamics in a tilted lattice model, where interaction and tilt frequencies coexist. By tuning the tilt to be weaker than, yet comparable to, the interaction energy, we identify a tilt-mode CR facilitated by tunneling between lattice sites. Remarkably, the amplitudes of the two modes are controlled by the tunneling-to-interaction ratio prior to the quench. With further analysis and appropriate rescaling, we find that the two-mode amplitudes exhibit universal linear behavior when rescaled to their crossing point, independent of the tilt. These findings are supported by numerical simulations.

\begin{figure*}
	\centering
	\includegraphics[width=\textwidth]{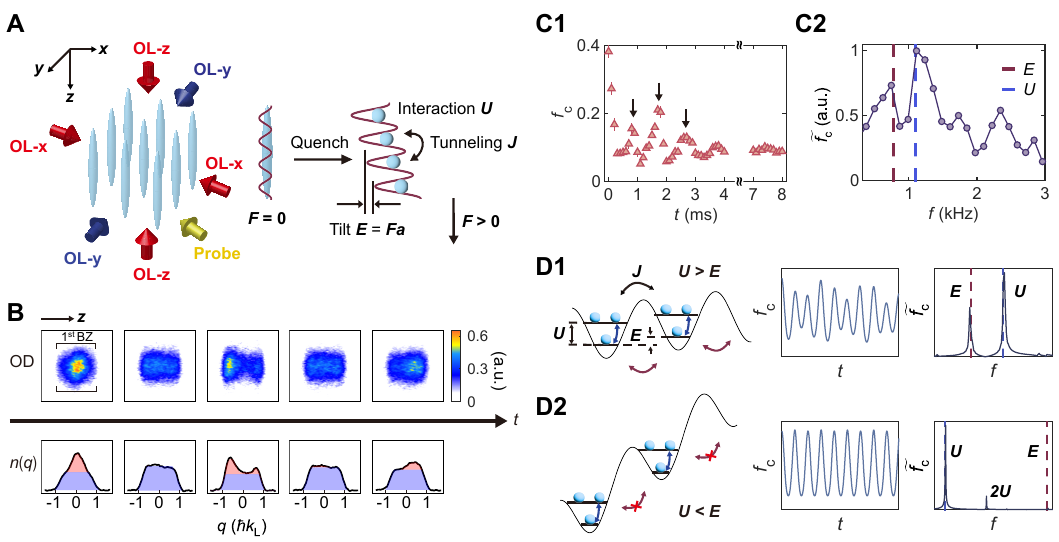}
    \caption{\textbf{Experimental setup and measurements of collapse and revival dynamics.} 
    (\textbf{A}) Schematic of the experimental system. A 3D cubic optical lattice is formed by three sets of lattice beams: two red-detuned beams ($\lambda = 1064~\mathrm{nm}$) along the $x$ and $z$ axes, and one blue-detuned beam ($\lambda' = 760~\mathrm{nm}$) along the $y$ axis. The probe beam intersects at an angle of $45^{\circ}$ with respect to the $x$ axis in the $x$-$y$ plane. The 1D bosonic gas is prepared with zero net force. By suddenly reducing the magnetic levitation along the $z$ axis, a tilt quench is introduced, triggering the dynamics, while the lattice depth along the tube is either increased or held constant.
    (\textbf{B}) Measured quasi-momentum distribution $n(q)$ as a function of evolution time, illustrating the collapse and revival (CR) dynamics. The distributions in the lower panel are obtained by integrating the optical density (OD) shown in the upper panel. The coherent and incoherent components are highlighted by the red and blue shaded regions, respectively.
    (\textbf{C}) Experimental measurement of the coherent fraction $\fc$ (\textbf{C1}) and its spectrum $\widetilde{f_c}$ (\textbf{C2}) for $E = 772\ \mathrm{Hz}$, with parameters $J_i = 133\ \mathrm{Hz}$, $U_i = 810\ \mathrm{Hz}$, $J_f = 20\ \mathrm{Hz}$, and $U_f = 1105\ \mathrm{Hz}$. The first three revival peaks in \textbf{(C1)} are marked by arrows. The solid line in \textbf{(C2)} serves as a guide to the eye.
    (\textbf{D}) Numerical simulations of CR dynamics for a no-trap system ($\omega = 0$) in the regimes \textbf{(D1)} $U > E$ ($U_f = 1.23E$) and \textbf{(D2)} $U < E$ ($U_f = 0.29E$), showing the corresponding time evolution of $\fc$ and spectra $\widetilde{\fc}$. The parameters $J_{i,f}$ and $U_{i,f}$ are identical to those in \textbf{(C)}. The red and blue dashed lines in \textbf{(C2)} and \textbf{(D)} indicate $E$ and $U$, respectively. a.u., arbitrary units.
    }
	\label{fig1}
\end{figure*}

\begin{figure*}
	\centering
    \includegraphics[width=\textwidth]{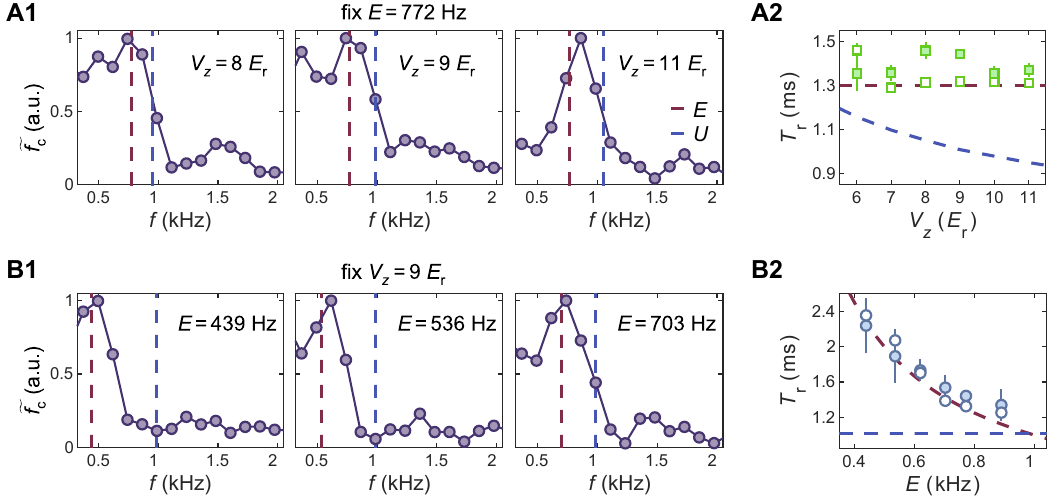}
	\caption{\textbf{Evidence of tilt-mode ($E$-mode) CR .} 
	(\textbf{A1}) Spectra of coherent fraction $\widetilde{\fc}$ for $V_z = 8,\ 9,$ and $11\ \Er$ with $E = 772\ \textrm{Hz}$. $V_z$ is kept fixed after the quench of $E$. (\textbf{A2}) CR period $T_{\rm r}$ as a function of $V_z$ (green squares). (\textbf{B1}) $\widetilde{\fc}$ for $E = 439,\ 536,$ and $703\ \textrm{Hz}$, with $V_z = 9\ \Er$ ($J=49\rm\ Hz$, $U = 987\rm\ Hz$) held constant following the quench of tilt. (\textbf{B2}) CR period $T_{\rm r}$ as a function of tilt $E$ (blue circles). In both \textbf{A} and \textbf{B}, the red and blue dashed lines denote the $E$ and $U$, respectively. The CR period is extracted by fitting a sum of Gaussian functions (S3), with error bars obtained using the bootstrap method. The white data points indicate the numerical simulation results for $T_{\rm r}$, analyzed with the same fitting procedure as in the experiment. The solid lines in the spectra serve as a guide to the eye. a.u., arbitrary units.}
	\label{fig2}
\end{figure*}
\section*{RESULTS}
\noindent {\bf Implementing tilted lattice systems for CR dynamics}\\
Our model is based on an array of one-dimensional (1D) ultracold bosons subjected to a tilted potential in an optical lattice. The 1D bosons are first prepared in the 3D lattice, as demonstrated in Fig.~\ref{fig1}A. After suddenly applying a constant force $F$ to the 1D bosons while ramping up the lattice, the system is expected to exhibit collapse and revival of coherence. The physics can be captured by the 1D Bose-Hubbard (BH) Hamiltonian in the presence of a tilt and a harmonic trap,
\begin{align}
	\hat{H}_{\rm 1D}=&-J\sum_{\langle j,l\rangle}\hat{a}^{\dagger}_j\hat{a}_l+\frac{U}{2}\sum_{j}\hat{n}_j(\hat{n}_j-1)\nonumber\\
	&+E\sum_{j}j\hat{n}_j+V_{\rm T}\sum_{j}j^2\hat{n}_j,
	\label{eq1}
\end{align}
where $\hat{a}^{\dagger}_j$ ($\hat{a}_j$) are the boson creation (annihilation) operators at site $j$, $\hat{n}_j=\hat{a}^{\dagger}_j\hat{a}_j$ are the number operators, $J$ is the nearest-neighbor tunneling strength, $U$ is the two-body interaction energy, $E=Fa$ is the potential energy difference between two adjacent lattice sites, with $a$ the lattice spacing, and $V_{\rm T}$ is the harmonic potential energy. Notably, a quench of the lattice depth along the 1D direction modifies both $U$ and $J$, whereas a quench of the tilt affects only the linear gradient term. Later in this paper, we denote the parameters before the quench as $J_i$, $U_i$, and $E_i = 0$, and those after the quench as $J_f$, $U_f$, and $E_f = E$.

In our experiment, we start with a BEC of typically $8\times 10^4$ $^{87}\rm Rb$ atoms initially trapped in the crossed-beam optical dipole trap and levitated against gravity by a magnetic field gradient. A 3D cubic optical lattice (OL) is constituted by two red-detuned lattice beams (with wavelength $\lambda = 2a= 1064$ nm) along $x-z$ plane (OL-$x$ and OL-$z$) and one blue-detuned lattice beam ($\lambda' = 760$ nm) along $y$ axis (OL-$y$), as shown in Fig.~\ref{fig1}A. To load the BEC adiabatically into the lowest orbital ($S$ band) of the lattice, the three lattice beams are ramped up within 80 ms and held for 20 ms, while the dipole trap is gradually turned off. The lattice depths for OL-$x$ and OL-$y$ throughout all the measurements are $V_x=35\ \Er$, $V_y=39\ \Er$,  where $\Er=h^2k_{\mathrm{L}}^2/2m=2\ \mathrm{kHz}$ is the recoil energy, with $k_{\textrm{L}}=2\pi/\lambda$ the wave vector. The deep 2D lattices OL-$x$ and OL-$y$ create an array of decoupled 1D tubes with an average filling $\bar{n}\sim2$ per lattice site along the $z$ axis, ensuring that tunneling in the $x-y$ plane remains suppressed throughout the experimental time scale. After the loading process, the initial lattice depth of OL-$z$ $V^i_{z}$ is either maintained or quenched to a specific value $V^f_{z}$, while the tilted potential $E$ is applied by lowering the magnetic levitation. This process simultaneously initiates the CR and BO dynamics. 

Notably, in previous works mentioned above, CR dynamics is explored in the regime $U_f<E$ and solely governed by on-site interactions ~\cite{PhysRevA.89.023606,PhysRevLett.90.213002, PhysRevLett.112.193003}. 
In our current setup, given the $s$-wave scattering length of $^{87}\mathrm{Rb}$ atoms $a_s \approx 5.3\ \textrm{nm}$, the on-site interaction $U$ of our experiment ranges from 810 Hz to 1105 Hz for $V_z$ between $5$ and $13~E_{\mathrm{r}}$. Meanwhile, the tilt energy $E$ can be tuned from 450 Hz to 900 Hz. This parameter range helps us access the new regime $U_f\gtrsim E$ and investigate whether new modes related to inter-site processes emerge.

\noindent {\bf Measurement of CR dynamics}\\
The main quantity we study for the CR behaviors is the coherent fraction $\fc$, which quantifies the coherence of the system. In the experiment, we apply the designed band mapping scheme to measure $\fc$~\cite{PhysRevLett.127.200601,Liang2022}. In practice, we switch off the OL-$x$ and OL-$y$ components immediately, while the OL-$z$ component is ramped down exponentially within 1 ms, mapping the quasi-momentum space to the momentum space. This scheme allows for the accurate measurement of the quasi-momentum distribution $n(q)$ in an interacting system by non-adiabatically releasing the interaction energy, thus preventing distortion in the measurements caused by interactions during the time-of-flight (TOF) process.

Fig.~\ref{fig1}B shows the quasi-momentum distribution $n(q)$ measured via the band mapping scheme during a typical CR process. Initially, a coherent peak appears at $q = 0$. After some evolution, the atoms dephase and spread nearly uniformly across the first Brillouin zone (BZ), with coherence later recovering as the system undergoes both CR and BO. The distribution is given by $n(q)=n_{\rm coh}(q)+n_{\rm incoh}(q)$, with $n_{\rm coh}(q)$ (red shaded area) the coherent (condensed) atoms forming peaks, and $n_{\rm incoh}(q)$ (blue shaded area) the incoherent (non-condensed) atoms forming a plateau. The coherent fraction is then calculated as $\fc = \int \mathrm{d}q\ n_{\rm coh}(q)/N$~\cite{PhysRevLett.127.200601,Liang2022}, where $N$ is the total atom number (see more details in Supplementary material~S1).

The typical time evolution of $\fc$ is displayed in Fig.~\ref{fig1}C1 for a quench of the lattice depth from $V^i_{z} = 5\ \Er$ to $V^f_{z} = 13\ \Er$ together with a quench of tilt $E = 772\ \rm Hz$. The evolution exhibits CR dynamics with three clearly resolved revivals (see three peaks in Fig.~\ref{fig1}C1, marked by arrows) and a slow dephasing, with a lifetime of approximately 3.0 ms. The reason for this gradual dephasing is mainly attributed to the harmonic trap, with a trapping frequency of $\omega = 2\pi\times20.2(3)\rm\ Hz$ along the tubes in our system (see S2 for details). The spectrum for $\fc$ is analyzed using Fourier transformation, named as $\widetilde{\fc}$, as shown in Fig.~\ref{fig1}C2. Aside from observing a peak corresponding to the two-body interaction (indicated by blue dashed line) in $\widetilde{\fc}$, another characteristic frequency corresponding to the tilt (marked by red dashed line) is identified. The CR dynamics manifest as a two-frequency oscillatory behavior. For simplicity, we denote the interaction- and tilt-mode CR as the $U$ and $E$ modes, respectively.

The two-mode oscillatory behavior in $\fc$ is benchmarked by the simulation for a no-trap system ($\omega=0$) using density matrix renormalization group (DMRG) and time-evolving block decimation (TEBD) methods (see S2 for details), as shown in Fig.~\ref{fig1}D1. For comparison, we simulate another system with $U_f<E$ (Fig.~\ref{fig1}D2), where $\fc(t)$ exhibits a nearly single-mode oscillation associated with $U_f$, consistent with Ref.~\cite{PhysRevLett.90.213002,PhysRevA.89.023606,PhysRevLett.112.193003}.

\noindent {\bf Evidence of tilt-mode CR}\\
To provide further experimental evidence for the $E$-mode CR, we perform a first set of measurements without quenching the lattice depth, i.e., $V^i_{z}=V^f_{z}=V_z$. The results are shown in Fig.~\ref{fig2}. First, we measure $\fc$ and the corresponding spectra $\widetilde{\fc}$ for various $V_z$ with a quench of tilt $E = 772\ \rm Hz$. As shown in Fig.~\ref{fig2}A1, all $\widetilde{\fc}$ spectra peak near $f = E/h$ (red dashed line), indicating the dominance of the $E$ mode. In this case, we fit the time-domain $\fc$ with a series of Gaussian peaks and compute the CR period $T_\textrm{r}$ as the average spacing between adjacent peaks~\cite{Greiner2002,PhysRevA.99.013602} (see S3 for details). We find that $T_\textrm{r} = 1.40(5)\ \rm ms$ remains constant across different $V_z$ (Fig.~\ref{fig2}A2), consistent with the $E$-mode period (red dashed line) and deviating from the $U$-mode period (blue dashed line). Second, we measure $\widetilde{\fc}$ for varying tilt $E$, keeping $V_z = 9\ \Er$ ($J = 49\ \rm Hz$, $U = 987\ \rm Hz$). The peaks in $\widetilde{\fc}$ shift in response to different tilt values (Fig.~\ref{fig2}B1). Using the same fitting procedure, the extracted CR periods scale inversely with $E$, consistent with $T_\textrm{r} = h/E$ (Fig.~\ref{fig2}B2). These results provide direct experimental evidence of $E$-mode CR. The revival periods obtained from TEBD simulations including the harmonic trap (white data points in Fig.~\ref{fig2}) agree with the experimental results within 9.9\%.

The emergence of the $E$ mode originates from the interplay between tilt and interaction in the many-body energy spectrum, which naturally separates into the regimes $U_f < E$ and $U_f > E$. We do not consider the case $U_f \approx E$, where resonant tunneling occurs~\cite{Meinert2014}. When $U_f < E$, the atoms in individual sites are effectively independent, as their energy levels are predominantly shifted by the tilt. Consequently, the 1D bosonic system is split into decoupled sites following the quench of $E$, with atoms at each site undergoing CR at a frequency set by $U_f$ (Fig.~\ref{fig1}D2). In contrast, for $U_f > E$, neighboring sites remain coupled through tunneling after the quench. This gives rise to $E$-mode CR, in which tunneling mediates the phase evolving at frequency $E/h$ from one site to its neighbors (Fig.~\ref{fig1}D1). This picture is supported by simulations with $J_f = 0$, where the $E$ mode vanishes while the $U$ mode persists, as shown in S2, S4, and Ref.~\cite{PhysRevA.89.023606}. While the $U$-mode CR reflects on-site dynamics, the $E$-mode CR can be interpreted as an inter-site process facilitated by tunneling.

\begin{figure}
	\centering
	\includegraphics[width=1\columnwidth]{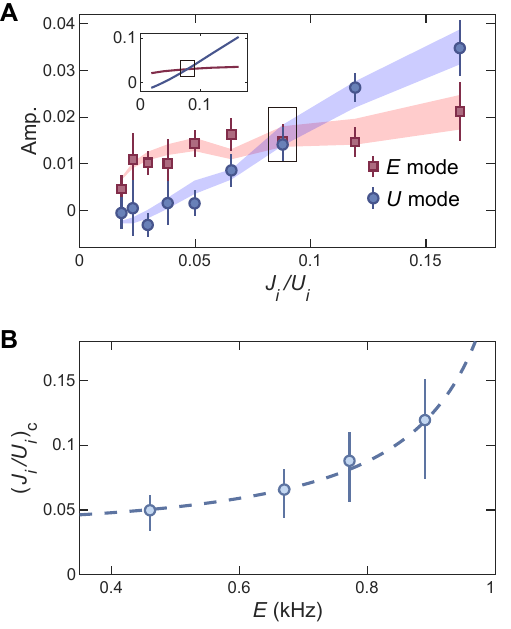}
	\caption{\textbf{Crossover of two-mode CR amplitudes.} 
		(\textbf{A}) Amplitudes of the $E$-mode (red squares) and $U$-mode (blue circles) CR as a function of $J_i/U_i$ for $E = 772\ \textrm{Hz}$, $J_f = 20\ \textrm{Hz}$, and $U_f = 1105\ \textrm{Hz}$. Error bars are evaluated using the bootstrap method. The red and blue shaded regions denote the numerically predicted amplitudes of the $E$ and $U$ modes, respectively, with uncertainties obtained using the same fitting procedure as in the experiment. Inset: numerical simulation results for a homogeneous system. The black boxes indicate the crossing point at $(J_i/U_i)_{\textrm{c}} = 0.08$. (\textbf{B}) Crossing points in the initial tunneling-to-interaction ratio for the amplitudes of the two modes at different tilts $E$. Error bars arise from the resolution of $V_{z,i}$ in the measurements. The dashed line denotes the predictions from numerical simulations for a homogeneous system.}
	\label{fig3}
\end{figure}

\noindent {\bf Crossover of two-mode amplitudes}\\
In a second set of measurements, we investigate the more general case where the $U$ and $E$ modes coexist for $U_f > E$. By introducing a finite quench of the lattice depth in this regime, the $U$-mode oscillation is activated in the CR dynamics. This naturally raises the question of how the strengths of the two modes depend on the parameters $(J_{i,f}, U_{i,f}, E)$. Specifically, we examine how the two-mode amplitudes vary with $J_i/U_i$ for $V^i_{z}$ ranging from 5 to 13 $\Er$ at $E = 772\ \textrm{Hz}$, while keeping the final lattice depth fixed at $V^f_{z} = 13\ \Er$ ($J_f = 20\ \textrm{Hz}$, $U_f = 1105\ \textrm{Hz}$). We extract the amplitudes of these two modes by fitting the data to a combination of cosine functions with an offset:
\begin{align}
	\fc(t) = A_E \cos\left(\frac{Et}{\hbar}\right) 
	+ A_{U} \cos\left(\frac{U_f t}{\hbar}\right) + B,
	\label{eq2}
\end{align}
where $A_U$ and $A_E$ are the amplitudes of the $U$ and $E$ modes, respectively. Due to the finite lifetime of $\fc(t)$ in the experiment, we limit the fitting process to the time window from $t=0$ to $t=3$ ms. The detail of extracting $A_U$ and $A_E$ by using bootstrap is presented in S3.

The $U$-mode (blue circles) and $E$-mode (red squares) amplitudes as functions of $J_i/U_i$ are shown in Fig.~\ref{fig3}A. A crossover between $U$- and $E$-mode amplitudes is observed, which manifests as a transition in the CR dynamics from $U$-mode dominance to $E$-mode dominance. These results are confirmed by TEBD simulations that include the harmonic trap, as indicated by the red and blue shaded areas. Moreover, our simulation for the no-trap system also displays the same crossing point at $(J_i/U_i)_{\textrm{c}}=0.08$ (inset of Fig.~\ref{fig3}A), suggesting that the transition of the dominating mode is less affected by the harmonic trap in the system.

To quantitatively analyze the dependence of the amplitudes on the BH parameters, we propose a simple analytic model based on perturbation theory that yields $\fc$. In general, the $E$-mode oscillation of $\fc$ arises from interference in the off-diagonal elements of $g^{(1)}(j,l)$. For sufficiently small tunneling, only nearest-neighbor correlations contribute significantly to these off-diagonal terms. Therefore, the evolution of $\fc$ is primarily governed by the phase of $g^{(1)}(j,j\pm1)$. Further details of the model are provided in S4. From this model, $A_U$ and $A_E$ are given by
\begin{align}
	A_E =  C'_{E}\frac{J_f}{\Er}\frac{J_i}{U_i} + \frac{C_{U}}{2}\frac{J_f}{U_f-E} + \frac{C_{U}}{2}\frac{J_f}{U_f+E},
	\label{eq3}
\end{align}
\begin{align}
	A_{U} = C_{U}\frac{J_i}{U_i} 
	- \frac{C_{U}}{2}\frac{J_f}{U_f-E} - \frac{C_{U}}{2}\frac{J_f}{U_f+E},
	\label{eq4}
\end{align}
where $C'_E$ and $C_{U}$ are coefficients independent of $J_{i,f}$, $U_{i,f}$, and $E$. The first term in $A_E$ arises from second-order perturbation. This model is valid only in the limit $J_i,\ J_f \ll E,\ |U_f-E|,\ U_f+E,\ U_i$. Notably, the amplitudes remain unchanged under inversion of the tilt $E \to -E$.

In the following, we elucidate the origin of the crossover between the two-mode amplitudes in Fig.~\ref{fig3}A. Based on numerical simulation results for a homogeneous system (see S1), the coefficient of $J_i/U_i$ in $A_E$ is $C'_E J_f/\Er = 0.100(5)$ (Eq.~\eqref{eq3}). This value is much smaller than the corresponding coefficient in $A_U$, for which $C_U = 0.760(11)$ (Eq.~\eqref{eq4}). The difference in slopes arises because increasing $J_i/U_i$ enhances on-site atom-number fluctuations~\cite{Orzel2001,Greiner2002_SFMI,PhysRevLett.98.200405,Sherson2010}, favoring the $U$-mode CR. In contrast, inter-site dynamics remain unaffected by these fluctuations to first order. At lower $J_i/U_i$, the suppression of on-site number fluctuations further reduces $A_U$. Consequently, the two modes respond differently to $J_i/U_i$, leading to a crossover of their amplitudes.

\begin{figure}
	\centering
	\includegraphics[width=1\columnwidth]{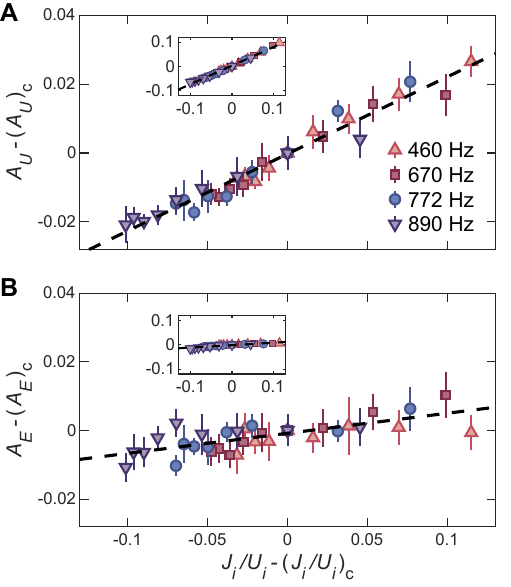}
    \caption{\textbf{Universal linear scaling of the two modes.} 
    Rescaled amplitudes of the $U$ mode (\textbf{A}) and $E$ mode (\textbf{B}) around the crossing point for various tilt values $E$. The black dashed lines indicate linear fits to the data. Error bars are obtained using the bootstrap method. Inset: numerical simulation results for a homogeneous system.}
	\label{fig4}
\end{figure}

To further investigate how the tilt affects the crossing point, we repeat the above measurements for various tilt. Fig.~\ref{fig3}B shows the crossing point of $(J_i/U_i)_{\textrm{c}}$ at different tilt values. As the tilt increases, the crossing point shifts to a higher $J_i/U_i$. According to our analytic model, the crossing point of the two modes writes
\begin{align}
	\left(\frac{J_i}{U_i}\right)_{\textrm{c}}
	=\frac{C_U}{C_U-C_{E}}\left(\frac{J_f}{U_f-E}+\frac{J_f}{U_f+E}\right),
	\label{eq5}
\end{align}
which follows a hyperbolic-like dependence on $E$, with $C_E=C'_E J_f/\Er$. Since $E$ and $U_f$ are in the same magnitude, the $J_f/(U_f-E)$ term, which is a dimensionless parameter describing the resonant behavior as mentioned in Ref.~\cite{PhysRevA.70.015604}, dominates the crossing point due to its much larger contribution. Moreover, the crossing point is mainly controlled by the factor $C_U/(C_U - C_E)$. We plot the analytic curve (dashed line) for the crossing point in Fig.~\ref{fig3}B based on Eq.~\eqref{eq5} for a homogeneous system. This curve aligns with our data in the trapped system within the error bars, suggesting that the crossing point is less affected by the trap.

\noindent {\bf Universal linear scaling of two-mode amplitudes}\\
Finally, we demonstrate the universality of these two modes around the crossing point. As indicated by the model, both amplitudes vary linearly with $J_i/U_i$. Upon rescaling to their crossing point, they are expected to show the universal linear scaling under fixed $J_f$ and $U_f$:
\begin{align}
	A_{U(E)}-\left(A_{U(E)}\right)_{\textrm{c}}=C_{U(E)}\left[\,\frac{J_i}{U_i}-\left(\frac{J_i}{U_i}\right)_{\textrm{c}}\,\right],
	\label{eq6}
\end{align}
which is independent of $E$. We plot the amplitudes of the $U$ and $E$ modes as a function of $J_i/U_i - (J_i/U_i)_{\textrm{c}}$ in Fig.~\ref{fig4}. The data collapse onto two distinct linear trends (black dashed lines). Linear fits to the data yield slopes of $C_U = 0.224(8)$ and $C_E = 0.059(9)$ for the $U$ and $E$ modes, respectively. The slopes for the trapped system are smaller than those for the no-trap system. In general, trap-induced dephasing dampens both the $U$-mode and $E$-mode oscillations of $\fc$, leading to a more gradual dependence of the amplitudes on $J_i/U_i$. For comparison with the measurements, we perform numerical simulations for systems without a trap. The simulation results are presented in the inset of Fig.~\ref{fig4}. For the no-trap system, the data still exhibit universal linear behavior after rescaling to the crossing point. Thus, the linear scaling behavior is robust against the trap. The analytic model for a homogeneous system already captures most features of the dynamics.

\section*{DISCUSSION}
Summarizing, we have observed the two-mode collapse and revival behavior in an ensemble of 1D bosons with the presence of a tilted optical lattice, by accessing a new parameter regime where the interaction is stronger than and of comparable magnitude to the tilt. When both the lattice depth and the tilt are rapidly increased, the system exhibits a combination of tilt- and interaction-mode oscillations, which is captured in the coherent fraction. Moreover, we show that the rescaled amplitudes of the two oscillations exhibit universal linear scaling for a fixed final state, independent of tilt. 
These findings provide direct evidence for the new mechanism of atomic dephasing and revival, leading to a more profound understanding of phase coherence in many-body correlated systems.

The effects of higher-order interactions in tilted lattice systems may become accessible by improving the homogeneity of the system~\cite{PhysRevA.89.023606,Will2010}, which would further extend the lifetime of the dynamics. In the present setup, although residual harmonic confinement along the 1D tubes is not completely canceled, this could be reduced by introducing additional blue-detuned laser beams. Moreover, the observed tilt-mode CR may be connected to other dynamical processes in tilted lattice systems. For example, doublon formation in systems with unit filling~\cite{Simon2011,PhysRevLett.111.053003} may occur together with similar coherence dynamics. For other atomic species with available Feshbach resonances and lighter atomic masses (e.g., $^{7}\textrm{Li}$), such phenomena may further extend to the strongly interacting regime. Such platforms provide promising directions for extending the present observations.

\section*{ACKNOWLEDGMENTS}
The authors thank Xiong-Jun Liu and Hanns-Christoph N\"agerl for their helpful discussions. The DMRG and TEBD calculations were performed using the TeNPy Library~\cite{tenpy}.

\noindent {\bf Funding:}
This work is supported by the National Natural Science Foundation of China (Grants No. 92365208), National Key Research and Development Program of China (Grants No. 2021YFA0718300 and No. 2021YFA1400900), and The Fundamental Research Funds for the Central Universities, Peking University.

\noindent {\bf Author contributions:}
C.-K. L., T. Z., H. Y. and X. Z. conceived this work. C.-K. L., S. J. and Z. H. designed the experiments and performed the measurements. C.-K. L., S. J. and F. W. analyzed the data. C.-K. L., Y. H. and C. L. performed the numeric simulations. Y. H. developed the analytical model. H. Y. and X. Z. supervised this work. All authors read, edited, and approved the final manuscript.

\noindent {\bf Competing interests:}
There are no competing interests to declare.

\noindent {\bf Data and materials availability:}
The data shown in this manuscript is available via Zenodo~\cite{zenodo2026}.

\section*{Supplementary Materials}
\noindent {\bf The pdf file includes}:\\
Supplementary Text S1 to S4\\
Figs. S1 to S9\\
References

\bibliographystyle{revtex}
\bibliography{my.bib}

@article{sherson2010,
  title = {Single-atom-resolved fluorescence imaging of an atomic {M}ott insulator},
  author = {Jacob F. Sherson and Christof Weitenberg and Manuel Endres and Marc Cheneau and Immanuel Bloch and Stefan Kuhr},
  journal = {\Jnature},
  volume = {467},
  pages = {68--72},
  year = {2010}
}

@article{greiner2002,
  title={Quantum phase transition from a superfluid to a {M}ott insulator in a gas of ultracold atoms}, 
  volume={415},
  number={6867},
  journal={\Jnature},
  publisher={Nature Publishing Group},
  author={Greiner, Markus and Mandel, Olaf and Esslinger, Tilman and H\"ansch, Theodor W and Bloch, Immanuel},
  year={2002},
  pages={39--44}
}

@article{trotzky2008,
  title={Time-resolved observation and control of superexchange interactions with ultracold atoms in optical lattices},
  volume={319},
  number={5861},
  journal={\Jscience},
  publisher={AAAS},
  author={Trotzky, S and Cheinet, P and F\"olling, S and Feld, M and Schnorrberger, U and Rey, A M and Polkovnikov, A and Demler, E A and Lukin, Mikhail D and Bloch, I},
  year={2008},
  pages={295--299}
}

@article{PhysRev.108.1175,
  title = {Theory of Superconductivity},
  author = {Bardeen, J. and Cooper, L. N. and Schrieffer, J. R.},
  journal = {Phys. Rev.},
  volume = {108},
  issue = {5},
  pages = {1175--1204},
  numpages = {0},
  year = {1957},
  publisher = {American Physical Society}
}

@article{PhysRev.131.2766,
	title = {Coherent and Incoherent States of the Radiation Field},
	author = {Glauber, Roy J.},
	journal = {Phys. Rev.},
	volume = {131},
	issue = {6},
	pages = {2766--2788},
	numpages = {0},
	year = {1963},
	publisher = {American Physical Society}
}

@article{PhysRev.140.A1051,
	title = {Stimulated Emission of Radiation in a Single Mode},
	author = {Cummings, F. W.},
	journal = {Phys. Rev.},
	volume = {140},
	issue = {4A},
	pages = {A1051--A1056},
	numpages = {0},
	year = {1965},
	publisher = {American Physical Society}
}

@article{PhysRevLett.44.1323,
	title = {Periodic Spontaneous Collapse and Revival in a Simple Quantum Model},
	author = {Eberly, J. H. and Narozhny, N. B. and Sanchez-Mondragon, J. J.},
	journal = {Phys. Rev. Lett.},
	volume = {44},
	issue = {20},
	pages = {1323--1326},
	numpages = {0},
	year = {1980},
	publisher = {American Physical Society}
}

@book{klauder1985coherent,
  title={Coherent states: applications in physics and mathematical physics},
  author={Klauder, John R and Skagerstam, Bo-Sture},
  year={1985},
  publisher={World scientific}
}

@article{PhysRevLett.58.353,
	title = {Observation of quantum collapse and revival in a one-atom maser},
	author = {Rempe, Gerhard and Walther, Herbert and Klein, Norbert},
	journal = {Phys. Rev. Lett.},
	volume = {58},
	issue = {4},
	pages = {353--356},
	numpages = {0},
	year = {1987},
	publisher = {American Physical Society}
}

@article{RevModPhys.62.867,
  title = {Coherent states: Theory and some applications},
  author = {Zhang, Wei-Min and Feng, Da Hsuan and Gilmore, Robert},
  journal = {Rev. Mod. Phys.},
  volume = {62},
  issue = {4},
  pages = {867--927},
  numpages = {0},
  year = {1990},
  publisher = {American Physical Society}
}

@article{PhysRevLett.76.1800,
	title = {Quantum Rabi Oscillation: A Direct Test of Field Quantization in a Cavity},
	author = {Brune, M. and Schmidt-Kaler, F. and Maali, A. and Dreyer, J. and Hagley, E. and Raimond, J. M. and Haroche, S.},
	journal = {Phys. Rev. Lett.},
	volume = {76},
	issue = {11},
	pages = {1800--1803},
	numpages = {0},
	year = {1996},
	publisher = {American Physical Society}
}

@article{PhysRevLett.76.4508,
	title = {Bloch Oscillations of Atoms in an Optical Potential},
	author = {Ben Dahan, Maxime and Peik, Ekkehard and Reichel, Jakob and Castin, Yvan and Salomon, Christophe},
	journal = {Phys. Rev. Lett.},
	volume = {76},
	issue = {24},
	pages = {4508--4511},
	numpages = {0},
	year = {1996},
	publisher = {American Physical Society}
}

@article{PhysRevLett.77.2158,
	title = {Collapses and Revivals of Bose-Einstein Condensates Formed in Small Atomic Samples},
	author = {Wright, E. M. and Walls, D. F. and Garrison, J. C.},
	journal = {Phys. Rev. Lett.},
	volume = {77},
	issue = {11},
	pages = {2158--2161},
	numpages = {0},
	year = {1996},
	publisher = {American Physical Society}
}

@article{PhysRevLett.78.2511,
	title = {Inhibition of Coherence in Trapped Bose-Einstein Condensates},
	author = {Imamo\ifmmode \bar{g}\else \={g}\fi{}lu, A. and Lewenstein, M. and You, L.},
	journal = {Phys. Rev. Lett.},
	volume = {78},
	issue = {13},
	pages = {2511--2514},
	numpages = {0},
	year = {1997},
	publisher = {American Physical Society}
}

@article{andrews1997twoBEC,
author = {M. R. Andrews  and C. G. Townsend  and H.-J. Miesner  and D. S. Durfee  and D. M. Kurn  and W. Ketterle },
title = {Observation of Interference Between Two Bose Condensates},
journal = {Science},
volume = {275},
number = {5300},
pages = {637-641},
year = {1997}
}

@article{Orzel2001,   author = {Orzel, C. and Tuchman, A. K. and Fenselau, M. L. and Yasuda, M. and Kasevich, M. A.},   title = {Squeezed States in a Bose-Einstein Condensate},   journal = {Science},   volume = {291},   number = {5512},   pages = {2386-2389},   note = {doi: 10.1126/science.1058149},   year = {2001},   type = {Journal Article}}

@article{Greiner2002_SFMI,   author = {Greiner, Markus and Mandel, Olaf and Esslinger, Tilman and H{\"a}nsch, Theodor W. and Bloch, Immanuel},   title = {Quantum phase transition from a superfluid to a Mott insulator in a gas of ultracold atoms},   journal = {Nature},   volume = {415},   number = {6867},   pages = {39-44},   ISSN = {1476-4687}, year = {2002}, type = {Journal Article}}

@article{PhysRevLett.90.213002,
	title = {New Bloch Period for Interacting Cold Atoms in 1D Optical Lattices},
	author = {Kolovsky, Andrey R.},
	journal = {Phys. Rev. Lett.},
	volume = {90},
	issue = {21},
	pages = {213002},
	numpages = {4},
	year = {2003},
	publisher = {American Physical Society}
}

@article{PhysRevA.70.015604,
	title = {Bloch oscillations in the Mott-insulator regime},
	author = {Kolovsky, Andrey R.},
	journal = {Phys. Rev. A},
	volume = {70},
	issue = {1},
	pages = {015604},
	numpages = {4},
	year = {2004},
	publisher = {American Physical Society}
}

@article{PhysRevLett.97.060402,
	title = {Long-Lived Bloch Oscillations with Bosonic Sr Atoms and Application to Gravity Measurement at the Micrometer Scale},
	author = {Ferrari, G. and Poli, N. and Sorrentino, F. and Tino, G. M.},
	journal = {Phys. Rev. Lett.},
	volume = {97},
	issue = {6},
	pages = {060402},
	numpages = {4},
	year = {2006},
	publisher = {American Physical Society}
}

@article{PhysRevA.74.053616,
	title = {Hard-core bosons on optical superlattices: Dynamics and relaxation in the superfluid and insulating regimes},
	author = {Rigol, Marcos and Muramatsu, Alejandro and Olshanii, Maxim},
	journal = {Phys. Rev. A},
	volume = {74},
	issue = {5},
	pages = {053616},
	numpages = {13},
	year = {2006},
	publisher = {American Physical Society}
}

@article{PhysRevLett.98.180601,
	title = {Quench Dynamics and Nonequilibrium Phase Diagram of the Bose-Hubbard Model},
	author = {Kollath, Corinna and L\"auchli, Andreas M. and Altman, Ehud},
	journal = {Phys. Rev. Lett.},
	volume = {98},
	issue = {18},
	pages = {180601},
	numpages = {4},
	year = {2007},
	publisher = {American Physical Society}
}

@article{PhysRevLett.98.200405,
	title = {Preparing and Probing Atomic Number States with an Atom Interferometer},
	author = {Sebby-Strabley, J. and Brown, B. L. and Anderlini, M. and Lee, P. J. and Phillips, W. D. and Porto, J. V. and Johnson, P. R.},
	journal = {Phys. Rev. Lett.},
	volume = {98},
	issue = {20},
	pages = {200405},
	numpages = {4},
	year = {2007},
	publisher = {American Physical Society}
}

@article{Johnson_2009,
	doi = {10.1088/1367-2630/11/9/093022},
	url = {https://doi.org/10.1088/1367-2630/11/9/093022},
	year = {2009},
	publisher = {},
	volume = {11},
	number = {9},
	pages = {093022},
	author = {Johnson, P R and Tiesinga, E and Porto, J V and Williams, C J},
	title = {Effective three-body interactions of neutral bosons in optical lattices},
	journal = {New Journal of Physics}
}

@article{Will2010,
	author={Will, Sebastian
	and Best, Thorsten
	and Schneider, Ulrich
	and Hackerm{\"u}ller, Lucia
	and L{\"u}hmann, Dirk-S{\"o}ren
	and Bloch, Immanuel},
	title={Time-resolved observation of coherent multi-body interactions in quantum phase revivals},
	journal={Nature},
	year={2010},
	volume={465},
	number={7295},
	pages={197-201},
	issn={1476-4687}
}

@article{Simon2011,   author = {Simon, Jonathan and Bakr, Waseem S. and Ma, Ruichao and Tai, M. Eric and Preiss, Philipp M. and Greiner, Markus},   title = {Quantum simulation of antiferromagnetic spin chains in an optical lattice},   journal = {Nature},   volume = {472},   number = {7343},   pages = {307-312},   ISSN = {1476-4687}, year = {2011}, type = {Journal Article}}

@article{PhysRevA.83.063609,
	title = {Collapse and revival dynamics of number-squeezed superfluids of ultracold atoms in optical lattices},
	author = {Tiesinga, E. and Johnson, P. R.},
	journal = {Phys. Rev. A},
	volume = {83},
	issue = {6},
	pages = {063609},
	numpages = {6},
	year = {2011},
	publisher = {American Physical Society}
}

@article{2013VlastakisCatState,
author = {Brian Vlastakis  and Gerhard Kirchmair  and Zaki Leghtas  and Simon E. Nigg  and Luigi Frunzio  and S. M. Girvin  and Mazyar Mirrahimi  and M. H. Devoret  and R. J. Schoelkopf },
title = {Deterministically Encoding Quantum Information Using 100-Photon Schr\"odinger Cat States},
journal = {Science},
volume = {342},
number = {6158},
pages = {607-610},
year = {2013}
}

@article{PhysRevLett.111.053003,
	title = {Quantum Quench in an Atomic One-Dimensional Ising Chain},
	author = {Meinert, F. and Mark, M. J. and Kirilov, E. and Lauber, K. and Weinmann, P. and Daley, A. J. and N\"agerl, H.-C.},
	journal = {Phys. Rev. Lett.},
	volume = {111},
	issue = {5},
	pages = {053003},
	numpages = {5},
	year = {2013},
	publisher = {American Physical Society}
}

@article{PhysRevA.89.023606,
	title = {Bloch oscillations and quench dynamics of interacting bosons in an optical lattice},
	author = {Mahmud, K. W. and Jiang, L. and Tiesinga, E. and Johnson, P. R.},
	journal = {Phys. Rev. A},
	volume = {89},
	issue = {2},
	pages = {023606},
	numpages = {12},
	year = {2014},
	publisher = {American Physical Society}
}

@article{PhysRevLett.112.193003,
	title = {Interaction-Induced Quantum Phase Revivals and Evidence for the Transition to the Quantum Chaotic Regime in 1D Atomic Bloch Oscillations},
	author = {Meinert, F. and Mark, M. J. and Kirilov, E. and Lauber, K. and Weinmann, P. and Gr\"obner, M. and N\"agerl, H.-C.},
	journal = {Phys. Rev. Lett.},
	volume = {112},
	number = {19},
	pages = {193003},
	numpages = {5},
	year = {2014},
	doi = {10.1103/PhysRevLett.112.193003},
	publisher = {American Physical Society}
}

@article{Meinert2014,
	author={Meinert, Florian
	and Mark, Manfred J.
	and Kirilov, Emil
	and Lauber, Katharina
	and Weinmann, Philipp
	and Gr{\"o}bner, Michael
	and Daley, Andrew J.
	and N{\"a}gerl, Hanns-Christoph},
	title={Observation of many-body dynamics in long-range tunneling after a quantum quench},
	journal={Science},
	year={2014},
	publisher={American Association for the Advancement of Science},
	volume={344},
	number={6189},
	pages={1259-1262}
}

@book{bickel2015mathematical,
  title={Mathematical statistics: basic ideas and selected topics, volumes I-II package},
  author={Bickel, Peter J and Doksum, Kjell A},
  year={2015},
  publisher={Chapman and Hall/CRC}
}

@article{PhysRevLett.120.213201,
	title = {Observation and Uses of Position-Space Bloch Oscillations in an Ultracold Gas},
	author = {Geiger, Zachary A. and Fujiwara, Kurt M. and Singh, Kevin and Senaratne, Ruwan and Rajagopal, Shankari V. and Lipatov, Mikhail and Shimasaki, Toshihiko and Driben, Rodislav and Konotop, Vladimir V. and Meier, Torsten and Weld, David M.},
	journal = {Phys. Rev. Lett.},
	volume = {120},
	issue = {21},
	pages = {213201},
	numpages = {6},
	year = {2018},
	publisher = {American Physical Society}
}

@article{PhysRevA.99.013602,
	title = {Observation of atom-number fluctuations in optical lattices via quantum collapse and revival dynamics},
	author = {Zhou, Tianwei and Yang, Kaixiang and Zhu, Zijie and Yu, Xudong and Yang, Shifeng and Xiong, Wei and Zhou, Xiaoji and Chen, Xuzong and Li, Chen and Schmiedmayer, J\"org and Yue, Xuguang and Zhai, Yueyang},
	journal = {Phys. Rev. A},
	volume = {99},
	issue = {1},
	pages = {013602},
	numpages = {6},
	year = {2019},
	publisher = {American Physical Society}
}

@article{GUO20222291,
	title = {Quantum precision measurement of two-dimensional forces with 10-28-Newton stability},
	journal = {Science Bulletin},
	volume = {67},
	number = {22},
	pages = {2291-2297},
	year = {2022},
	issn = {2095-9273},
	author = {Xinxin Guo and Zhongcheng Yu and Fansu Wei and Shengjie Jin and Xuzong Chen and Xiaopeng Li and Xibo Zhang and Xiaoji Zhou}
}

@article{PhysRevLett.124.043204,
	title = {Enhanced Superexchange in a Tilted Mott Insulator},
	author = {Dimitrova, Ivana and Jepsen, Niklas and Buyskikh, Anton and Venegas-Gomez, Araceli and Amato-Grill, Jesse and Daley, Andrew and Ketterle, Wolfgang},
	journal = {Phys. Rev. Lett.},
	volume = {124},
	issue = {4},
	pages = {043204},
	numpages = {6},
	year = {2020},
	publisher = {American Physical Society}
}

@article{PhysRevLett.127.200601,
	title = {Observation of Many-Body Quantum Phase Transitions beyond the Kibble-Zurek Mechanism},
	author = {Huang, Qi and Yao, Ruixiao and Liang, Libo and Wang, Shuai and Zheng, Qinpei and Li, Dingping and Xiong, Wei and Zhou, Xiaoji and Chen, Wenlan and Chen, Xuzong and Hu, Jiazhong},
	journal = {Phys. Rev. Lett.},
	volume = {127},
	issue = {20},
	pages = {200601},
	numpages = {6},
	year = {2021},
	publisher = {American Physical Society}
}

@article{Alexander2022,
	author = {Alexander Aeppli  and Anjun Chu  and Tobias Bothwell  and Colin J. Kennedy  and Dhruv Kedar  and Peiru He  and Ana Maria Rey  and Jun Ye },
	title = {Hamiltonian engineering of spin-orbit-coupled fermions in a Wannier-Stark optical lattice clock},
	journal = {Science Advances},
	volume = {8},
	number = {41},
	pages = {eadc9242},
	year = {2022}
}

@article{Liang2022,
	author={Liang, Libo
	and Zheng, Wei
	and Yao, Ruixiao
	and Zheng, Qinpei
	and Yao, Zhiyuan
	and Zhou, Tian-Gang
	and Huang, Qi
	and Zhang, Zhongchi
	and Ye, Jilai
	and Zhou, Xiaoji
	and Chen, Xuzong
	and Chen, Wenlan
	and Zhai, Hui
	and Hu, Jiazhong},
	title={Probing quantum many-body correlations by universal ramping dynamics},
	journal={Science Bulletin},
	year={2022},
	volume={67},
	number={24},
	pages={2550-2556},
	issn={2095-9273}
}

@article{Huang2023,
	title = {Measurement of interacting quantum phases: A band mapping scheme},
	journal = {Frontiers of Physics},
	volume = {18},
	pages = {52307-},
	year = {2023},
	issn = {2095-0462},
	doi = {https://doi.org/10.1007/s11467-023-1326-y},
	url = {https://journal.hep.com.cn/fop/EN/10.1007/s11467-023-1326-y},
	author = {Huang, Qi and Zhu, Zijie and Wang, Yifei and Liang, Libo and Zheng, Qinpei and Chen, Xuzong}

}

@article{PhysRevResearch.5.013136,
	title = {Dimensional crossover of quantum critical dynamics in many-body phase transitions},
	author = {Zheng, Qinpei and Wang, Yuqing and Liang, Libo and Huang, Qi and Wang, Shuai and Xiong, Wei and Zhou, Xiaoji and Chen, Wenlan and Chen, Xuzong and Hu, Jiazhong},
	journal = {Phys. Rev. Res.},
	volume = {5},
	issue = {1},
	pages = {013136},
	numpages = {9},
	year = {2023},
	publisher = {American Physical Society},
}

@article{tirole2023double,
  title={Double-slit time diffraction at optical frequencies},
  author={Tirole, Romain and Vezzoli, Stefano and Galiffi, Emanuele and Robertson, Iain and Maurice, Dries and Tilmann, Benjamin and Maier, Stefan A and Pendry, John B and Sapienza, Riccardo},
  journal={Nature Physics},
  volume={19},
  number={7},
  pages={999--1002},
  year={2023},
  publisher={Nature Publishing Group UK London}
}

@article{Leonard2023,   author = {L{\'e}onard, Julian and Kim, Sooshin and Kwan, Joyce and Segura, Perrin and Grusdt, Fabian and Repellin, Cécile and Goldman, Nathan and Greiner, Markus},   title = {Realization of a fractional quantum Hall state with ultracold atoms},   journal = {Nature},   volume = {619},   number = {7970},   pages = {495-499}, ISSN = {1476-4687}, url = {https://doi.org/10.1038/s41586-023-06122-4}, year = {2023}, type = {Journal Article}}

@article{
	Zhu2024,
	author = {Zijie Zhu  and Marius G\"achter  and Anne-Sophie Walter  and Konrad Viebahn  and Tilman Esslinger },
	title = {Reversal of quantized Hall drifts at noninteracting and interacting topological boundaries},
	journal = {Science},
	volume = {384},
	number = {6693},
	pages = {317-320},
	year = {2024},
	doi = {10.1126/science.adg3848}
}

@article{Wang:24,
	author = {Yuqing Wang and Libo Liang and Qinpei Zheng and Qi Huang and Wenlan Chen and Jing Zhang and Xuzong Chen and Jiazhong Hu},
	journal = {Opt. Express},
	number = {23},
	pages = {41657--41664},
	publisher = {Optica Publishing Group},
	title = {Divergence of thermalization rates driven by the competition between finite temperature and quantum coherence},
	volume = {32},
	year = {2024}
}

@article{Dhar2025,
	author = {Dhar, Sudipta and Wang, Botao and Horvath, Milena and Vashisht, Amit and Zeng, Yi and Zvonarev, Mikhail B. and Goldman, Nathan and Guo, Yanliang and Landini, Manuele and N\"agerl, Hanns-Christoph},
	title = {Observing anyonization of bosons in a quantum gas},
	journal = {Nature},
	volume = {642},
	number = {8066},
	pages = {53-57},
	ISSN = {1476-4687},
	year = {2025},
	type = {Journal Article}
}

@article{2np8-mp39,
	title = {Anyonization of Bosons in One Dimension: An Effective Swap Model},
	author = {Wang, Botao and Vashisht, Amit and Guo, Yanliang and Dhar, Sudipta and Landini, Manuele and N\"agerl, Hanns-Christoph and Goldman, Nathan},
	journal = {Phys. Rev. Lett.},
	volume = {135},
	issue = {25},
	pages = {253403},
	numpages = {8},
	year = {2025},
	publisher = {American Physical Society},
	doi = {10.1103/2np8-mp39},
	url = {https://link.aps.org/doi/10.1103/2np8-mp39}
}

@article{tenpy,
	title={{Efficient numerical simulations with Tensor Networks: Tensor Network Python (TeNPy)}},
	author={Johannes Hauschild and Frank Pollmann},
	journal={SciPost Phys. Lect. Notes},
	pages={5},
	year={2018},
	publisher={SciPost}
}

@article{penrose1956,
  title = {Bose-Einstein Condensation and Liquid Helium},
  author = {Penrose, Oliver and Onsager, Lars},
  journal = {Phys. Rev.},
  volume = {104},
  issue = {3},
  pages = {576--584},
  numpages = {0},
  year = {1956},
  publisher = {American Physical Society}
}

@book{strang2016linear,
  title     = {Introduction to Linear Algebra},
  author    = {Strang, Gilbert},
  edition   = {5th},
  year      = {2016},
  publisher = {Wellesley-Cambridge Press},
  isbn      = {9780980232776}
}

@article{gray2006toeplitz,
  title   = {Toeplitz and Circulant Matrices: A Review},
  author  = {Gray, Robert M.},
  journal = {Foundations and Trends in Communications and Information Theory},
  volume  = {2},
  number  = {3},
  pages   = {155--239},
  year    = {2006},
  doi     = {10.1561/0100000006}
}

@article{2026AcousticWavePlasmon,
author = {Skyler P. Selvin  and Majid Esfandyarpour  and Anqi Ji  and Yan Joe Lee  and Colin Yule  and Jung-Hwan Song  and Mohammad Taghinejad  and Mark L. Brongersma },
title = {Acoustic wave modulation of gap plasmon cavities},
journal = {Science},
volume = {389},
number = {6759},
pages = {516-520},
year = {2025},
doi = {10.1126/science.adv1728}
}

@article{2020_AcousticWaveNonreciprocity,
author = {Piyush J. Shah  and Derek A. Bas  and Ivan Lisenkov  and Alexei Matyushov  and Nian X. Sun  and Michael R. Page },
title = {Giant nonreciprocity of surface acoustic waves enabled by the magnetoelastic interaction},
journal = {Science Advances},
volume = {6},
number = {49},
pages = {eabc5648},
year = {2020}
}

@article{wendt2026electrically,
  title={An electrically injected solid-state surface acoustic wave phonon laser},
  author={Wendt, Alexander and Storey, Matthew J and Miller, Michael and Anderson, Dalton and Chatterjee, Eric and Horrocks, William and Smith, Brandon and Wong, Ping-Show and Arterburn, Shawn and Friedmann, Thomas A and others},
  journal={Nature},
  volume={649},
  number={8097},
  pages={597--603},
  year={2026},
  publisher={Nature Publishing Group UK London}
}

@article{rabec2025bloch,
  title={Bloch oscillations of a soliton in a one-dimensional quantum fluid},
  author={Rabec, F and Chauveau, G and Brochier, G and Nascimbene, S and Dalibard, J and Beugnon, J},
  journal={Nature Physics},
  volume={21},
  number={10},
  pages={1541--1547},
  year={2025},
  publisher={Nature Publishing Group UK London}
}

@article{zhou2021doubleslit,
author = {Haowen Zhou  and William E. Perreault  and Nandini Mukherjee  and Richard N. Zare },
title = {Quantum mechanical double slit for molecular scattering},
journal = {Science},
volume = {374},
number = {6570},
pages = {960-964},
year = {2021}
}

@article{hong2025fetosecond,
author = {EunHo Hong  and EunSeo Jang  and JunWoo Kim },
title = {Femtosecond spectroscopy with paired single photons: Emulating a double-slit experiment in the time-frequency domain},
journal = {Science Advances},
volume = {11},
number = {42},
pages = {eadw9759},
year = {2025}
}

@article{Chakraborti2025grapheneinterferometer,
author = {H. Chakraborti  and L. Pugliese  and A. Assouline  and K. Watanabe  and T. Taniguchi  and N. Kumada  and D. C. Glattli  and M. Jo  and H.-S. Sim  and P. Roulleau },
title = {Electron collision in a two-path graphene interferometer},
journal = {Science},
volume = {388},
number = {6746},
pages = {492-496},
year = {2025}
}

@article{pedalino2026probing,
  title={Probing quantum mechanics with nanoparticle matter-wave interferometry},
  author={Pedalino, Sebastian and Ram{\'\i}rez-Galindo, Bruno E and Ferstl, Richard and Hornberger, Klaus and Arndt, Markus and Gerlich, Stefan},
  journal={Nature},
  volume={649},
  number={8098},
  pages={866--870},
  year={2026},
  publisher={Nature Publishing Group UK London}
}

@article{greve2022entanglement,
  title={Entanglement-enhanced matter-wave interferometry in a high-finesse cavity},
  author={Greve, Graham P and Luo, Chengyi and Wu, Baochen and Thompson, James K},
  journal={Nature},
  volume={610},
  number={7932},
  pages={472--477},
  year={2022},
  publisher={Nature Publishing Group UK London}
}

@article{Li2026atomicsensors,
author = {Yifan Li  and Lex Joosten  and Youcef Baamara  and Paolo Colciaghi  and Alice Sinatra  and Philipp Treutlein  and Tilman Zibold },
title = {Multiparameter estimation with an array of entangled atomic sensors},
journal = {Science},
volume = {391},
number = {6783},
pages = {374-378},
year = {2026},
doi = {10.1126/science.adt2442}
}

@misc{zenodo2026,
  title        = {Data set is available from {Zenodo} at doi: 10.5281/zenodo.20145701},
  publisher    = {Zenodo},
}

\renewcommand{\theequation}{S\arabic{equation}}
\setcounter{equation}{0}
\renewcommand{\thefigure}{S\arabic{figure}}
\setcounter{figure}{0}
\renewcommand{\thesection}{S\arabic{section}}
\setcounter{section}{0}
\onecolumngrid  


\newpage

{\centering \bf \large Supplemental Material for \\}
{\centering \bf \large
Two-mode collapse and revival of quantum coherent state in a tilted optical lattice\par \vspace{1cm}}

In this supplemental material, we provide details of the numerical simulations, the perturbation-theory description of the CR dynamics, and the fitting procedures used in the experiment. The structure of this supplemental material is as follows. In Sec.~\ref{S1}, we describe the procedure for extracting the coherent fraction $\fc$ from the measured quasi-momentum distribution. In Sec.~\ref{S2}, we outline the numerical simulations of the 1D bosonic chain, including the calculation of the coherent fraction and its time evolution. The methods used to determine the revival period $T_{\textrm{r}}$ and the two-mode amplitudes $A_{U(E)}$ from $\fc$ is presented in Sec.~\ref{S3}. Finally, in Sec.~\ref{S4},  we present the derivation of $\fc$ in terms of $(J_{i,f}, U_{i,f}, E)$ using perturbation theory.

\section{Evaluation of coherent fraction in the experiment}
\label{S1}
Using the band mapping scheme outlined in the main text of the manuscript, we measure the intact quasi-momentum distribution in the experiment, which reveals distinct peaks and plateaus corresponding to condensed (coherent) atoms and non-condensed (incoherent) atoms, respectively. The distributions of these two types of atoms are clearly distinguishable. Motivated by previous works~\cite{PhysRevLett.127.200601,Liang2022,Huang2023,PhysRevResearch.5.013136,Wang:24}, we therefore use two different fitting functions to describe their profiles.

For the condensed atoms, they exhibit a characteristic linear motion within the first Brillouin zone (1st BZ), in the context of Bloch oscillations, due to the applied external force. When these atoms reach the momentum values $ q = \pm \hbar k_{\mathrm{L}} $, they undergo Bragg reflection, leading to the appearance of two peaks in the distribution. This two-peak feature is clearly seen in Fig.~1(d) of the main text. To phenomenologically describe the profile of condensed (coherent) atoms, we use two inverted parabolic functions, which represent the symmetric peaks on either side of the Brillouin zone:
\begin{align}
	n_{\rm coh}(q) = A_{\rm l} \left[ 1-\left(\frac{q-q_{\rm l}}{\sigma_{\rm l}}\right)^2\right]^2 + A_{\rm r}\left[ 1-\left(\frac{q-q_{\rm r}}{\sigma_{\rm r}}\right)^2\right]^2,
	\label{coh}
\end{align}
where $ A_{\rm l(r)}$ are the amplitudes of the two clusters of condensed atoms located at $ q < 0 $ (or $ q > 0 $), $ q_{\rm l(r)} $ are their respective centers, and $ \sigma_{\rm l(r)} $ represent their widths. This model effectively captures the behavior of condensed atoms undergoing periodic motion and reflection in the Brillouin zone.

In contrast, the non-condensed (incoherent) atoms fill the first Brillouin zone and exhibit a trapezoidal-like distribution. This distribution is characteristic of incoherent atoms, which are not confined to specific quasi-momentum state. The profile of these incoherent atoms can be described by a piecewise function:
\begin{align}
	n_{\rm incoh}(q) = \begin{cases} 
		B & \quad (q < q_1), \\
		B + k(q - q_1) & \quad (q_1 \leq q < q_2), \\
		k(q_2 - q_1) + B & \quad (q_2 \leq q < q_3), \\
		\max\left[k(q_2 - q_1 - q + q_3), B \right] & \quad (q \geq q_3).
	\end{cases}
	\label{incoh}
\end{align}
Here, $ B $ represents the offset of the total distribution, $ k $ is the slope of the trapezoidal sides, and $ q_i $ are the coordinates of the vertices of the trapezoid. This function captures the shape of the incoherent atom distribution, which is spread across the entire 1st BZ.

The total quasi-momentum distribution $ n(q) $ is the sum of the coherent and incoherent components:
\begin{align}
	n(q) = n_{\rm coh}(q) + n_{\rm incoh}(q).
\end{align}
This bi-mode model effectively illustrates the periodic dephasing and revival dynamics of the system, where the coherent atoms undergo Bloch oscillations and the incoherent atoms contribute to the broad background. Finally, the coherent fraction $ \fc $ is evaluated by calculating the ratio of total condensed atoms to the total atom number in the system, i.e., $\fc=\int\mathrm{d}q\ n_{\rm coh}(q)/N$. This fraction provides a direct measure of the coherence in the system, reflecting the extent to which the atoms remain coherent.

\begin{figure}
	\centering
	\includegraphics[width=\textwidth]{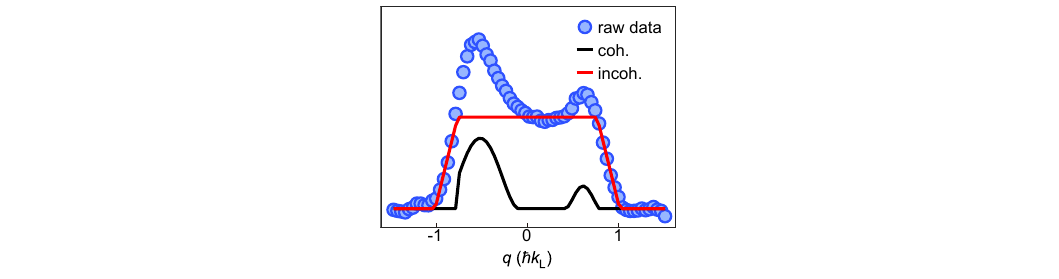}
	\caption{\textbf{Measured quasi-momentum distribution and fitting procedure.} 
		Measured quasi-momentum distribution $n(q)$ (points), together with the fitted coherent and incoherent components, $n_{\mathrm{coh}}(q)$ (red solid line) and $n_{\mathrm{incoh}}(q)$ (black solid line), obtained using Eqs.~\eqref{coh} and \eqref{incoh}.}
	\label{figS1}
\end{figure}

An example of the fitting procedure is shown in Fig.~\ref{figS1}. The coherent component exhibits the characteristic double-peak structure (black solid line), while the incoherent component forms a broad background (red solid line). The number of coherent (incoherent) atoms is obtained by integrating $n_{\rm coh}(q)$ ($n_{\rm incoh}(q)$) over quasi-momentum.

\section{Numeric simulation of the effective 1D system}
\label{S2}
In this work, we use the density matrix renormalization group (DMRG) method, implemented with the TeNPy library~\cite{tenpy}, to simulate the 1D bosonic chain. Within the DMRG framework, the many-body quantum state is represented as a matrix product state (MPS) $|\psi\rangle$. We first obtain the ground state of the initial system (before quenching the lattice depth or tilt) by minimizing the energy expectation value 
\begin{align}
	E = \frac{\langle\psi|\hat{H}_i|\psi\rangle}{\langle\psi|\psi\rangle},
\end{align}
where $\hat{H}_i$ denotes the initial Hamiltonian. Starting from this MPS ground state, we then compute the time evolution $|\psi(t)\rangle$ under the post-quench Hamiltonian $\hat{H}_f$ using the time-evolving block decimation (TEBD) method. In the simulations, we consider a effective 1D system of size $L=75$ with atom number $N_{\textrm{1D}}=100$, corresponding to an averaged tube in the experiment.

To characterize the coherence properties during the dynamics, we evaluate the single-particle correlation function
\begin{align}
	g^{(1)}(j,l;t) = \frac{\langle\psi(t)|\hat{a}_j^\dagger \hat{a}_l |\psi(t)\rangle}{\langle\psi|\psi\rangle},
\end{align}
which quantifies the degree of many-body coherence. The quasi-momentum distribution is obtained via Fourier transformation,
\begin{align}
	n(q;t) = \frac{1}{L}\sum_{j,l} e^{iq(r_j-r_l)} g^{(1)}(j,l;t),
\end{align}
where $q$ is the quasi-momentum. This provides a direct measure of phase coherence in quasi-momentum space.

To account for the harmonic confinement present in the experiment, we include an external trapping potential of the form
\begin{align}
	V_{\mathrm{T}} \sum_j j^2 \hat{n}j,
\end{align}
where $V_{\textrm{T}} = m\omega^2 a^2/2$, with $m$ the mass of $^{87}\mathrm{Rb}$, $\omega$ the trapping frequency, and $a$ the lattice spacing. In our experiment, the trapping frequency is $\omega_{\rm exp}=2\pi\times 20.2(3)\rm\ Hz$. The trapping potential is applied both before and after the quench. Its effect on the coherent dynamics is discussed below.

\subsection*{Calculating the coherent fraction in numeric simulations}
For the numerical simulations, there are two different definitions of coherent (condensate) fraction $\fc$. The first definition is given by
\begin{align}
	\fc=\frac{\lambda_{\textrm{max}}}{N},
	\label{1stdef}
\end{align}
where $\lambda_{\textrm{max}}$ is the largest eigenstate of matrix $g^{(1)}(j,l) = \langle \hat{a}^\dagger_j \hat{a}_l \rangle$, and $N$ is the total atom number. This definition is motivated by the Penrose--Onsager criterion~\cite{penrose1956}, which characterizes the coherent fraction as the ratio of the number of coherent atoms to the total atom number. The second definition is based on the quasi-momentum distribution $n(q)$, and is defined as~\cite{PhysRevA.89.023606}
\begin{align}
	\fc = \frac{n(q)_{\mathrm{max}}}{N},
	\label{2nddef}
\end{align}
where $n(q)_{\mathrm{max}}$ is the maximum value of $n(q)$.

\begin{figure}
	\centering
	\includegraphics[width=\textwidth]{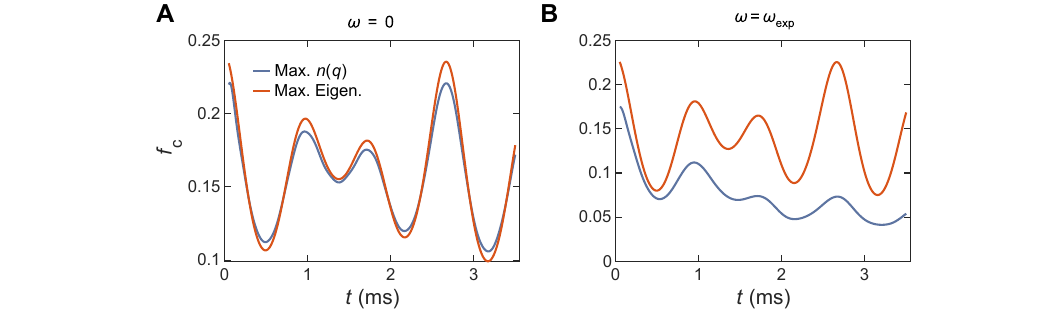}
	\caption{\textbf{Comparison of different definitions of the coherent fraction.} 
		Time evolution of the coherent fraction $\fc$ defined via the quasi-momentum distribution, $n(q)_{\max}/N$ (blue solid line), and via the largest eigenvalue of the single-particle correlation matrix, normalized by $N$ (orange solid line). The system is quenched from $J_i=80~\mathrm{Hz}$, $U_i=909~\mathrm{Hz}$, $E_i=0$ to $J_f=20~\mathrm{Hz}$, $U_f=1105~\mathrm{Hz}$, and $E=772~\mathrm{Hz}$. 
		\textbf{(A)} $\omega=0$. 
		\textbf{(B)} $\omega=\omega_{\rm exp}$.}
	\label{figS2}
\end{figure}

When $V_\textrm{T} = 0$, the system becomes spatially homogeneous and can be well described by periodic boundary conditions. In this case, the two definitions above are equivalent. This equivalence originates from the fact that, under periodic boundary conditions, the eigenstates of the system are momentum eigenstates. The single-particle correlation function $g^{(1)}(j,l)$ then takes the form of a circulant matrix, whose eigenvectors are plane-wave (Fourier) modes. Consequently, diagonalizing $g^{(1)}(j,l)$ in real space is equivalent to performing a Fourier transform to quasi-momentum space. For comparison, we perform simulations for a quench from $J_i=80\ \textrm{Hz},\ U_i=909\ \textrm{Hz},\ E_i=0$ to $J_f=20\ \textrm{Hz},\ U_f=1105\ \textrm{Hz},\ E=772\ \textrm{Hz}$ when $\omega=0$. We plot both $n(q)_{\rm max}/N$ and the maximum eigenvalue of $g^{(1)}(j,l)$ in Fig.~\ref{figS2}A, which show only a small deviation arising from the Fourier transformation procedure.

In the presence of harmonic confinement ($V_{\mathrm{T}}\neq 0$), however, the two definitions are no longer equivalent. The first definition remains formally well defined, as it directly probes the single-particle correlation matrix. Experimentally, by contrast, the coherent fraction is accessed via the quasi-momentum distribution, corresponding to the second definition. In this case, the eigenstates of the inhomogeneous system are projected onto quasi-momentum eigenstates, leading to a redistribution of weight over multiple quasi-momentum components. As a result, the extracted coherent fraction $\fc$ is systematically reduced compared to the Penrose-Onsager definition (see Fig.~\ref{figS2}B for $\omega=\omega_{\rm exp}$), and the quasi-momentum distribution exhibits finite broadening.

Although the first definition is insensitive to the trap, it is not experimentally accessible and does not capture the observed dephasing. Importantly, while the two definitions differ quantitatively for the trapping system, they capture the same dynamical behavior. For consistency with experimental observables, we therefore adopt the second definition of $\fc$ in Eq.~\eqref{2nddef} throughout our numerical simulations, both for $V_{\mathrm{T}}=0$ and $V_{\mathrm{T}}\neq 0$.

\subsection*{Effect of harmonic trap on CR dynamics}
As discussed in the main text and in previous works~\cite{Greiner2002,Will2010,PhysRevLett.112.193003}, the presence of a harmonic trap generally induces dephasing in coherent dynamics, where three to four revival peaks are typically observed in the experiment. 

To investigate the trapping effect, we perform numerical simulations over the time window $t\in[0,\,3.5]~\mathrm{ms}$ for a quench from $J_i=80~\mathrm{Hz}$, $U_i=909~\mathrm{Hz}$, $E_i=0$ to $J_f=20~\mathrm{Hz}$, $U_f=1105~\mathrm{Hz}$, and $E=772~\mathrm{Hz}$, with $\omega$ varied from $0$ to $2\pi\times60~\mathrm{Hz}$. 

\begin{figure}
	\centering
	\includegraphics[width=\textwidth]{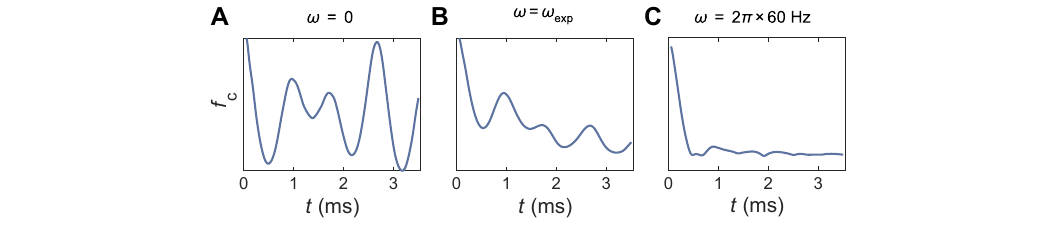}
	\caption{\textbf{Trap-induced dephasing of coherent revival dynamics.} 
		Time evolution of the coherent fraction $f_c$ for different trapping frequencies $\omega$: 
		\textbf{(A)} $0$, \textbf{(B)} $\omega_{\rm exp}$, and \textbf{(C)} $2\pi\times60~\mathrm{Hz}$. 
		The system is quenched from $J_i=80~\mathrm{Hz}$, $U_i=909~\mathrm{Hz}$, $E_i=0$ to $J_f=20~\mathrm{Hz}$, $U_f=1105~\mathrm{Hz}$, and $E=772~\mathrm{Hz}$.}
	\label{figS3}
\end{figure}

As shown in Fig.~\ref{figS3}A, three resolved revival peaks are observed for $\omega=0$. With increasing $\omega$, the amplitudes of these peaks gradually decrease, indicating gradual dephasing of the CR dynamics (see Fig.~\ref{figS3}B), which refers to the case for experiment. For $\omega= 2\pi\times60~\mathrm{Hz}$, the revival signatures are strongly suppressed (Fig.~\ref{figS3}C), showing that the CR dynamics is eventually obscured by the trapping effect.

\subsection*{Comparison between the cases of $J_f=0$ and $J_f\neq 0$}
In the experiment, the tunneling strength cannot be reduced to strictly zero, regardless of how deep the lattice potential along the tube is made. Numerical simulations, however, allow us to access the limiting case of $J_f=0$, thereby providing a complementary perspective for elucidating the underlying mechanism. To investigate the role of $J_f$ in the CR dynamics, particularly for the $E$ mode, we perform two sets of simulations corresponding to quenches from $J_i=80~\mathrm{Hz}$ ($U_i=909~\mathrm{Hz},\ E_i=0$) to either $J_f=20~\mathrm{Hz}$ or $J_f=0$ ($U_f=1105~\mathrm{Hz},\ E=772~\mathrm{Hz}$), while we set $\omega=0$.

\begin{figure}
	\centering
	\includegraphics[width=\textwidth]{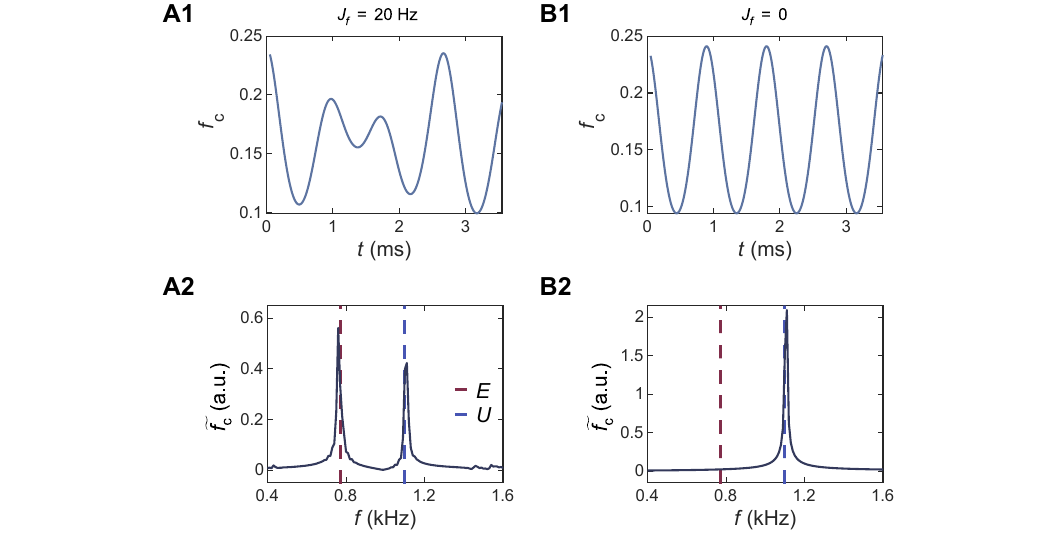}
	\caption{\textbf{Comparison of CR dynamics for $J_f=0$ and $J_f\neq0$.} 
		Time evolution of $\fc$ for \textbf{(A1)} $J_f=20~\mathrm{Hz}$ and \textbf{(A2)} $J_f=0$, together with the corresponding spectra $\widetilde{f_c}$ in \textbf{(B1)} and \textbf{(B2)}, respectively. The blue and red dashed lines indicate $E/h$ and $U/h$, respectively. The system is quenched from $J_i=80~\mathrm{Hz}$, $U_i=909~\mathrm{Hz}$, $E_i=0$ to the corresponding $J_f$, with $U_f=1105~\mathrm{Hz}$ and $E=772~\mathrm{Hz}$. The trapping frequency is $\omega=0$.}
	\label{figS4}
\end{figure}

For $J_f=20~\mathrm{Hz}$, the evolution of $\fc$ (Fig.~\ref{figS4}A1) exhibits a clear beating pattern arising from two characteristic frequencies, $E/h$ and $U/h$, as confirmed by the spectrum $\widetilde{\fc}$ shown in Fig.~\ref{figS4}B1. To achieve sufficient frequency resolution, the time evolution is computed up to $T=100~\mathrm{ms}$, corresponding to a spectral resolution $\delta f = 1/T = 10~\mathrm{Hz}$. This two-mode behavior has been discussed in the main text.

In contrast, for $J_f=0$, the dynamics of $\fc$ display a single-mode oscillation (Fig.~\ref{figS4}A2). Despite the presence of the tilt, the spectrum $\widetilde{\fc}$ exhibits only a single peak at $f=U/h$, with no signature of the $E$ mode (Fig.~\ref{figS4}B2). This can be understood from the fact that a finite tunneling $J_f$ is required to couple neighboring sites, enabling phase evolution associated with the tilt energy $E$ (i.e., $\propto e^{iEt/\hbar}$). When $J_f=0$, the inter-site coupling is absent, and the dynamics are governed solely by on-site interactions, leading to a pure $U$-mode CR without coexistence of the $E$ mode.

The results for $J_f=0$ are consistent with previous work~\cite{PhysRevA.89.023606}, as well as with the theoretical analysis presented in Sec.~\ref{S4}.

\section{Extraction of CR periods and two-mode amplitudes}
\label{S3}
In this section, we detail the procedures used to extract the coherent revival (CR) period $T_{\mathrm{r}}$ (Fig.~2) and the two-mode amplitudes $A_E$ and $A_U$ (Figs.~3 and 4). The uncertainties of these quantities are estimated using the bootstrap method~\cite{bickel2015mathematical}. For each dataset, synthetic datasets are generated by resampling the original data points with replacement and subsequently refitted using the same fitting procedure. Repeating this process 1000 times yields distributions of the fitted parameters, from which the uncertainties are obtained as the standard deviations. For the numerical results, the same fitting procedure is applied, with uncertainties determined from the fitting errors corresponding to the 95\% confidence interval.

\subsection*{Fitting CR periods with a sum of Gaussian peaks}
In Fig.~2A2 and B2, we present the revival period $T_{\textrm{r}}$ extracted from the time-domain coherent fraction $\fc$ without the quench of lattice depth. By solely applying the tilt quench, it exhibits dominant $E$-mode oscillation. As verified by the spectra in Fig.~2A1 and B1, the dynamics is characterized by a single dominant frequency $f=E/\hbar$. Therefore, we describe the revival sequence in $\fc$ using a sum of Gaussian peaks, following previous works~\cite{Greiner2002,PhysRevA.99.013602} with a slight modification:
\begin{align}
	\fc = A_0\exp(-t^2/\sigma_0^2) + \sum^N_{n=1} A_n \exp\!\left[-(t-T_{\textrm{r},n})^2/\sigma_n^2\right],
	\label{S10}
\end{align}
where $\sigma_n$ denotes the width of the $n$th peak and $T_{\textrm{r},n}$ its center. In the analysis, we typically include peaks up to $N\ge2$. As indicated by Eq.~\eqref{S10}, the widths of different peaks are allowed to vary. In addition, the peak centers approximately satisfy the relation 
$T_{\textrm{r},1} \approx T_{\textrm{r},2}/2 \approx T_{\textrm{r},3}/3 \approx \cdots$, 
although the spacing is not strictly equal. These are the two main features of the $E$-mode CR dynamics.

\begin{figure}
	\centering
	\includegraphics[width=\textwidth]{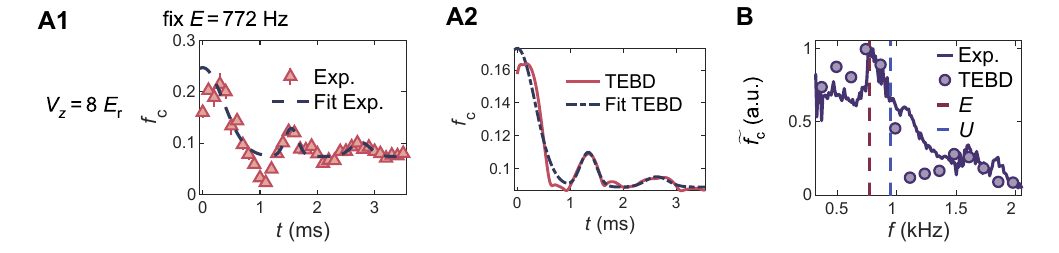}
	\caption{\textbf{Extraction of CR periods using a sum of Gaussian peaks.} 
		\textbf{(A)} Time evolution of $f_c$ at fixed $E=772~\mathrm{Hz}$, together with fits using Eq.~\eqref{S10}. The lattice depth is fixed at $V_z=8\ \Er$ after the tilt quench, while the trapping frequency is $\omega=\omega_{\rm exp}$. Experimental data (points) and numerical results (TEBD, solid lines) are shown in \textbf{(A1)} and \textbf{(A2)}, respectively. The black dashed and dash-dotted lines denote the corresponding fits to the experimental and numerical data. Error bars represent the standard errors from five measurements. 
		\textbf{(B)} Corresponding normalized spectra $\widetilde{f_c}$ for both experiment (points) and TEBD (purple solid line). The red and blue dashed lines indicate $E=772~\mathrm{Hz}$ and $U=950~\mathrm{Hz}$, respectively. a.u., arbitrary units.}
	\label{figS5}
\end{figure}

As an example, Fig.~\ref{figS5}A1,A2 shows $f_c$ at fixed $V_z=8~E_{\mathrm{r}}$ ($J=62.4~\mathrm{Hz}$, $U=950~\mathrm{Hz}$) following the tilt quench for both experiment and TEBD simulations. In the simulations, we set $\omega=\omega_{\rm exp}$. In both cases, the data are well described by Eq.~\eqref{S10}, as shown by the dashed (experiment) and dash-dotted (TEBD) lines. The corresponding normalized spectra are shown in Fig.~S5B and exhibit good agreement.

To determine the revival period, we evaluate the average period from the peak centers as
\begin{align}
	T_{\textrm{r}} = \frac{1}{N}\sum_{n=1}^{N}\frac{T_{\textrm{r},n}}{n},
	\label{S42}
\end{align}
where $N\ge2$ is the number of peaks included in the fitting. For the data shown in Fig.~\ref{figS7}, we take $N=2$, yielding $T_{\mathrm{r}}^{\mathrm{exp}}=1.460(38)\ \mathrm{ms}$ and $T_{\mathrm{r}}^{\mathrm{num}}=1.315(10)\ \mathrm{ms}$, which agree within $9.9\%$. This represents the maximum deviation across all measurements and simulations shown in Fig.~2.

\subsection*{Extraction of two-mode amplitudes using a sum of cosine functions}
Here, we describe the procedure used to extract the two-mode amplitudes $A_U$ and $A_E$ shown in Fig.~3. When a lattice quench is introduced, two-mode oscillations characterized by $U_f$ and $E$ can be observed in the regime $U_f>E$, as shown in Fig.~1C and D1. In the case where $\fc$ exhibits such two-mode oscillations, we fit the data using Eq.~(2). By fixing the frequencies $E$ and $U_f$ in the fitting function, we extract the corresponding amplitudes $A_U$ and $A_E$. Since the dynamics are accompanied by gradual dephasing, we restrict the fitting window to the first 3 ms of $\fc$ ($t=0\sim3\ \textrm{ms}$) to ensure consistent extraction of $A_U$ and $A_E$ for different values of $E$ and $J_i/U_i$.

\begin{figure}
	\centering
	\includegraphics[width=\textwidth]{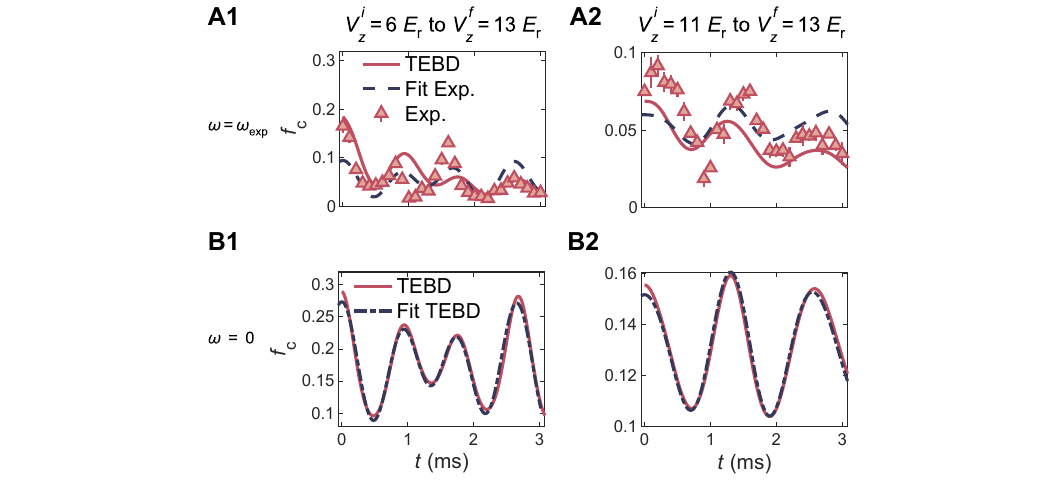}
	\caption{\textbf{Extraction of two-mode amplitudes using cosine fits.} 
		\textbf{(A)} Measured time evolution of $f_c$ at fixed $J_f=20~\mathrm{Hz}$, $U_f=1105~\mathrm{Hz}$, and $E=772~\mathrm{Hz}$, together with fits using Eq.~(2) and TEBD results. The initial lattice depths are (\textbf{A1}) $V_z^i=6~E_{\mathrm{r}}$ and (\textbf{A2}) $V_z^i=11~E_{\mathrm{r}}$. Error bars denote standard errors from five measurements. The dashed lines represent fits to the experimental data, while the solid lines denote TEBD results. 
		\textbf{(B)} Numerical simulations without harmonic confinement ($\omega=0$). The final-state parameters are the same as in (\textbf{A}), while the initial lattice depths are (\textbf{B1}) $V_z^i=6~E_{\mathrm{r}}$ and (\textbf{B2}) $V_z^i=11~E_{\mathrm{r}}$. The dash-dotted lines represent fits to the numerical data, and the solid lines denote TEBD results.}
	\label{figS6}
\end{figure}

As illustrative examples, Fig.~\ref{figS6}A1, A2 shows experimental data for quenches from $V_z^i=6~E_{\mathrm{r}}$ and $11~E_{\mathrm{r}}$ to $V_z^f=13~E_{\mathrm{r}}$, both at $E=772~\mathrm{Hz}$. The corresponding final-state parameters are $J_f=20~\mathrm{Hz}$ and $U_f=1105~\mathrm{Hz}$. The fits (dashed lines) agree well with the data, indicating that Eq.~(2) captures the dynamics. Within the fixed time window, three revival peaks are observed in Fig.~\ref{figS6}A1, corresponding to a $U$-mode-dominated regime ($A_U > A_E$), whereas only two revivals are visible in Fig.~\ref{figS6}A2, indicating an $E$-mode-dominated regime ($A_E > A_U$). This transition reflects the ratio $E/U \approx 0.7$. The experimental results are benchmarked against TEBD simulations (solid lines in Fig.~\ref{figS6}A1, A2).

For comparison, we also perform simulations without harmonic confinement ($\omega=0$) at $E=772~\mathrm{Hz}$, shown in Fig.~\ref{figS6}B. The same crossover from $U$-mode-dominated to $E$-mode-dominated dynamics is observed with increasing $V_z^i$, corresponding to decreasing $J_i/U_i$, consistent with the main text. In the absence of the trap, the agreement with Eq.~(2) is improved due to the absence of dephasing, as evidenced by the fits (dash-dotted lines). Although the experimental system exhibits slow trap-induced dephasing, this does not affect the crossover or the linear scaling. The extracted $A_U$ and $A_E$ for the homogeneous case are shown in the inset of Fig.~3.

In Fig.~4, we plot the rescaled amplitudes $A_{U(E)}-(A_{U(E)})_{\textrm{c}}$ as a function of $J_i/U_i-(J_i/U_i)_{\textrm{c}}$, where $(J_i/U_i)_{\textrm{c}}$ denotes the crossing point and $(A_U)_{\textrm{c}}=(A_E)_{\textrm{c}}$ the corresponding crossing amplitude. In practice, determining the crossing point precisely is challenging because we can only adjust $V^i_{z}$ rather than directly tune $U_i$, as the available Feshbach resonance of $^{87}\textrm{Rb}$ is not practical for controlling the interaction. Instead, we estimate $(J_i/U_i)_{\textrm{c}}$ by locating the minimum of $|A_U-A_E|$. The resolution of this estimate is limited by that of $V^i_{z}$, typically about $1\,\Er$.

\section{Time-dependent perturbation theory for the evolution of coherent fraction}
\label{S4}

In this section, we provide some theoretical discussions about the oscillations of the coherent fraction $\fc$ in the model where the Hamiltonian is 
\begin{align}
	\hat{H}_{i(f)}=&-J_{i(f)}\sum_{\langle j,l\rangle}\hat{a}^{\dagger}_j\hat{a}_l+\sum_{j}\frac{U_{i(f)}}{2}\hat{n}_j(\hat{n}_j-1)+E_{i(f)}\sum_{j}j\hat{n}_j
\end{align}
In this Hamiltonian, there is no $V_T$ term, i.e. we don't consider the presence of the harmonic trap. Because the harmonic trap potential is slowly varying and not involved in the mechanics we talk about here, which happens among a few neighboring sites. So according to \ref{S1} we use the largest eigenvalue definition of $\fc$.

We attribute the oscillation of $\fc$ to the interference of different frequency components in the off-diagonal elements of $g^{(1)}(j,l)$. Neglecting translation-symmetry breaking due to finite-size effects, $g^{(1)}(j,l)$ can be approximated as a Toeplitz matrix. Since correlations decay rapidly in the Mott regime, it is sufficient to retain only the nearest-neighbor terms $g^{(1)}(j,j+1)$. Assuming translation invariance, $g^{(1)}(j,j+1)=G^{(1)}(1)$, independent of $j$, so $\fc$ can be estimated as~\cite{strang2016linear,gray2006toeplitz}
\begin{equation}
	\fc=\lambda_{\max} \approx \bar{n}+2|g^{(1)}(j,j+1)|
	\label{eq:S2_eq1}
\end{equation}

In the following we discuss two cases: 1. tunneling $J_f=0$ (which is an ideal case discussed in \cite{PhysRevA.89.023606}) where only U-mode exists. 2. finite tunneling $J_f\neq0$ but is small and can be treated as perturbation term.

\subsection*{The case of $J_f=0$}
For $J_f=0$, we can actually demonstrate that there is only $U$-mode oscillation by "eliminating" $E$ term in Hamiltonian. To be specific, we perform a unitary transformation to eliminate the effect of Bloch oscillations:
\begin{align}
	\hat{U}(t)=\exp\!\Big(it\sum_j j E \hat{n}_j/\hbar\Big).
\end{align}
This shifts the potential term $E$ in the diagonal elements of Hamiltonian into the off-diagonal tunneling term, yielding $-J_f e^{iEt/\hbar}$ instead of $-J_f$. Consequently, when $J_f=0$, the parameter $E$ does not appear in the transformed Hamiltonian and influence the coherent fraction oscillation, contributing only to the Bloch oscillation of the momentum peak, while the $U$ mode remains. This procedure is equivalent to the similarity transformation
\begin{align}
	g^{(1)} \longrightarrow \Lambda g^{(1)} \Lambda^{-1}, \quad
	\Lambda=\mathrm{diag}(1,e^{iEt},e^{i2Et},\dots,e^{iLEt}),
\end{align}
which eliminates the phase accumulated from the $E$ potential while leaving the eigenvalues unchanged.

\subsection*{The case of finite tunneling $J_f\neq0$ while $J_f\ll U_f,\ U_f-E,\ E$}
\begin{figure}
	\centering
	\includegraphics[width=\textwidth]{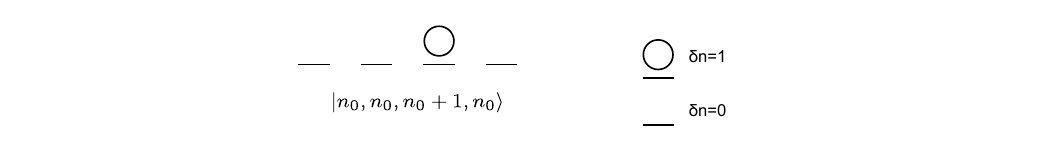}
	\caption{Toy model consisting of four sites used to explain the two-mode oscillation of $\fc$, where a Fock state of $|n_0,n_0,n_0+1,n_0\rangle$.}
	\label{figS7}
\end{figure}

For $J_f \neq 0$, a quantitative understanding of the emergence of the two-mode
oscillation in $\fc$ can be obtained by considering a simple toy model. In this model, only four sites of the full chain are retained, as illustrated in Fig.~\ref{figS7}. These four sites define a Hilbert subspace $\mathcal{H}_{\mathrm{sub}}$ of the full Hilbert space $\mathcal{H}$ of the one-dimensional system, such that
\[
\mathcal{H} = \mathcal{H}_{\mathrm{sub}} \oplus \mathcal{H}_{\mathrm{sub}}^{\perp}.
\]

Starting from the pure state of the full system, $\hat{\rho}_{\mathrm{tot}}(t) = |\Phi(t)\rangle\langle\Phi(t)|$,
the reduced density matrix within the subspace $\mathcal{H}_{\mathrm{sub}}$
is obtained by tracing out the complementary degrees of freedom,
\[
\hat{\rho}(t) =
\mathrm{Tr}_{\mathcal{H}_{\mathrm{sub}}^{\perp}}
\!\left(
|\Phi(t)\rangle\langle\Phi(t)|
\right).
\]

Within this subspace, we evaluate the single-particle correlation function
between the two central sites,
\[
g^{(1)}_{\mathrm{sub}}(2,3)
= \mathrm{Tr}_{\mathcal{H}_{\mathrm{sub}}}
\bigl(
\hat{\rho}(t)\,\hat{a}^{\dagger}_2 \hat{a}_3
\bigr),
\]
as an estimate of $g^{(1)}(i,i+1)=G^{(1)}(1)$, since the system is homogeneous and this local property can well represent the global property. 
This allows us to extract an estimate of the coherent fraction $\fc$
according to Eq.~\eqref{eq:S2_eq1}.
To proceed, we expand the reduced density matrix in the Fock basis,
\[
\hat{\rho}(t)
=
\sum_{\mathbf{n}, \mathbf{m}}
\rho_{\mathbf{n},\mathbf{m}}(t)
\exp\!\left(
\frac{i(E_{\mathbf{m}}-E_{\mathbf{n}})t}{\hbar}
\right)
|\mathbf{n}\rangle\langle\mathbf{m}|.
\]
So with this expansion, the correlation function becomes
\[
g^{(1)}_{\mathrm{sub}}(2,3)
=
\sum_{\mathbf{n}, \mathbf{m}}
\rho_{\mathbf{n},\mathbf{m}}(t)
\exp\!\left(
\frac{i(E_{\mathbf{m}}-E_{\mathbf{n}})t}{\hbar}
\right)
\langle\mathbf{m}|
\hat{a}^{\dagger}_2 \hat{a}_3
|\mathbf{n}\rangle.
\]
Therefore, the task reduces to evaluating all matrix elements
$\rho_{\mathbf{n},\mathbf{m}}(t)$ for which
$\langle\mathbf{m}|\hat{a}^{\dagger}_2 \hat{a}_3|\mathbf{n}\rangle \neq 0$.
This can be accomplished systematically using an order-by-order
perturbative expansion:$\rho_{\mathbf{n},\mathbf{m}}(0)$ can be obtained by stationary perturbation theory and $\rho_{\mathbf{n},\mathbf{m}}(t)$ can be obtained by time-dependent perturbation theory under the assumption that
$J_i/U_i,\;J_f/U_f,\;J_f/(U_f-E),\;J_f/(U_f+E) \ll 1$. In the following the derivations are presented in detail.

We first define the unperturbed Hamiltonian and its restriction to the four-site subspace(which contains only on-site terms) as
\[
\hat{H}^{(0)}
=
\frac{U}{2}\sum_j \hat{n}_j(\hat{n}_j-1)
+
E\sum_j j\hat{n}_j
,\;\;\;\;
\hat{H}^{(0)}_{\mathrm{sub}}
=
\frac{U}{2}\sum_{j=1}^{4}\hat{n}_j(\hat{n}_j-1)
+
E\sum_{j=1}^{4}j\hat{n}_j,
\]

The tunneling term and its subspace counterpart are treated as a perturbation term,
\[
\hat{H}^{(1)}
=
-J\sum_j \hat{a}^{\dagger}_j\hat{a}_{j+1} + h.c.
,\;\;\;\;
\hat{H}^{(1)}_{\mathrm{sub}}
=
-J\sum_{j=1}^{3}\hat{a}^{\dagger}_j\hat{a}_{j+1} + h.c.
\]

From our assumption, the dimensionless parameters $J_i/U_i,\;J_f/U_f,\;J_f/(U_f-E),\;J_f/(U_f+E) \ll 1$
are all small and of the same order, therefore only terms that are linear in these parameters are retained in the correlation function. 

The Fock states are labeled as
$|\mathbf{n}\rangle=|n_0+\delta n_1,n_0+\delta n_2,n_0+\delta n_3,n_0+\delta n_4\rangle=|\delta n_1,\delta n_2,\delta n_3,\delta n_4\rangle_{n_0}$,
where $n_0=\lfloor\bar{n}\rfloor$ denotes the integer part of the average filling and is fixed because of the homogeneity assumption. The reason of this labeling is that in our derivations $\delta n_j$ plays a key role, so our derivations can be universally applied to arbitrary $n_0$. Fig.~\ref{figS7} is an illustration of this.

Within a Hilbert subspace of fixed total particle-number deviation
$\delta n_{\mathrm{sum}}=\sum_{j=1}^4 \delta n_j$,
the low-energy subspace
$\mathcal{H}_{\mathrm{low}}\subset\mathcal{H}_{\mathrm{sub}}$
consists of states $|\mathbf{n}_L\rangle$ for which
$\delta n_j \in \{0,1\}$ and
$\rho^{(0)}_{\mathbf{n}_L,\mathbf{m}_L}\sim\mathcal{O}(1)$.
This follows from the fact that, for $\hat{H}^{(0)}$, the lowest-energy
configurations are obtained by starting from uniform filling with $n_0$
particles per site and distributing the remaining particles over the lattice.
Due to translational invariance, all such configurations are degenerate,
and the lowest-energy manifold is spanned by states with $\delta n_j=0,1$.

Because the initial tunneling $J_i$ is finite, linear contributions to the
correlation function arise from two distinct mechanisms.

Mechanism 1: linear terms can arise from the first-order stationary perturbation of the initial state. In this case, the ground-state density matrix acquires component $\lvert \mathbf{m}_H \rangle$ in the high-energy subspace, leading to couplings between states $\lvert \mathbf{m}_H \rangle$ and $\lvert \mathbf{n}_L \rangle$. Here, the state $\lvert \mathrm{m}_H \rangle$ is generated perturbatively from a low-energy state $\lvert \mathrm{q}_L \rangle$. As a result, the corresponding contribution to the correlation term $\rho_{\mathbf{n}_L,\mathbf{m}_H}(t)$ is 
\begin{align}
	\sum_{\mathbf{q}_L}
	\frac{J_i}{E^{i(0)}_{\mathbf{m}_H} - E^{i(0)}_{\mathbf{q}_L}} \, \langle\mathbf{q}_L|\sum_{j=1}^{3}\hat{a}^{\dagger}_j\hat{a}_{j+1}+h.c.|\mathbf{m}_H\rangle
	\langle\mathbf{m}_H|\hat a^{\dagger}_2\hat a_3 |\mathbf{n}_L\rangle
	\rho^{(0)}_{\mathbf{n}_L,\mathbf{q}_L}
	\mathrm{exp}\Big(\frac{i(E^{f(0)}_{\mathbf{m}_H}-E^{f(0)}_{\mathbf{n}_L})t}{\hbar}\Big),
	\label{eqS5}
\end{align}
where $E^{i,f(0)}_{\mathbf{m}_H}$ and $E^{i,f(0)}_{\mathbf{q}_L}$ denote the expectation values of the initial and final $\hat H^{(0)}_{sub}$ in the states $\lvert \mathbf{m}_H \rangle$ and $\lvert \mathbf{q}_L \rangle$ , and the corresponding contribution to the correlation term $\rho_{\mathbf{n}_H,\mathbf{m}_L}(t)$ is
\begin{align}
	\sum_{\mathbf{q}_L}
	\frac{J_i}{E^{i(0)}_{\mathbf{n}_H} - E^{i(0)}_{\mathbf{q}_L}} \, \langle\mathbf{n}_H|\sum_{j=1}^{3}\hat{a}^{\dagger}_j\hat{a}_{j+1}+h.c.|\mathbf{q}_L\rangle
	\langle\mathbf{m}_L|\hat a^{\dagger}_2\hat a_3 |\mathbf{n}_H\rangle
	\rho^{(0)}_{\mathbf{q}_L,\mathbf{m}_L}
	\mathrm{exp}\Big(\frac{i(E^{f(0)}_{\mathbf{m}_L}-E^{f(0)}_{\mathbf{n}_H})t}{\hbar}\Big),
	\label{eqS6}
\end{align}

Mechanism 2: linear contributions originate from the first-order time-dependent perturbation induced by the post-quench tunneling $J_f$. In this case, a low-energy state $\lvert \mathrm{q}_L \rangle$ is coupled to other states $\lvert \mathrm{m} \rangle$, which subsequently couple to $\lvert \mathrm{n}_L \rangle$. The resulting contribution to the correlation term $\rho_{\mathbf{n}_L,\mathbf{m}}(t)$ is
\begin{align}
	&\sum_{\mathbf{q}_L}
	\frac{J_f}{E^{f(0)}_{\mathbf{m}} - E^{f(0)}_{\mathbf{q}_L}} \, \langle\mathbf{q}_L|\sum_{j=1}^{3}\hat{a}^{\dagger}_j\hat{a}_{j+1}+h.c.|\mathbf{m}\rangle
	\langle\mathbf{m}|\hat a^{\dagger}_2\hat a_3 |\mathbf{n}_L\rangle
	\rho^{(0)}_{\mathbf{n}_L,\mathbf{q}_L}\nonumber\\
	&\times\Big(\mathrm{exp}\Big(\frac{i(E^{f(0)}_{\mathbf{q}_L}-E^{f(0)}_{\mathbf{n}_L})t}{\hbar}\Big)-\mathrm{exp}\Big(\frac{i(E^{f(0)}_{\mathbf{m}}-E^{f(0)}_{\mathbf{n}_L})t}{\hbar}\Big)\Big),
	\label{eqS7}
\end{align}
with  $E^{i,f(0)}_{\mathbf{m}_H}$ and $E^{i,f(0)}_{\mathbf{q}_L}$ again referring to the corresponding expectation values of the initial and final $\hat H^{(0)}_{sub}$ in the states $\lvert \mathbf{m}_H \rangle$ and $\lvert \mathbf{q}_L \rangle$. Similarly, the resulting contribution to the correlation term $\rho_{\mathbf{n},\mathbf{m}_L}(t)$ is
\begin{align}
	&\sum_{\mathbf{q}_L}
	\frac{J_f}{E^{f(0)}_{\mathbf{n}} - E^{f(0)}_{\mathbf{q}_L}} \, \langle\mathbf{n}|\sum_{j=1}^{3}\hat{a}^{\dagger}_j\hat{a}_{j+1}+h.c.|\mathbf{q}_L\rangle
	\langle\mathbf{m}_L|\hat a^{\dagger}_2\hat a_3 |\mathbf{n}\rangle
	\rho^{(0)}_{\mathbf{q}_L,\mathbf{m}_L}\nonumber\\
	&\times\Big(\mathrm{exp}\Big(\frac{i(E^{f(0)}_{\mathbf{m}_L}-E^{f(0)}_{\mathbf{q}_L})t}{\hbar})-\mathrm{exp}(\frac{i(E^{f(0)}_{\mathbf{m}_L}-E^{f(0)}_{\mathbf{n}})t}{\hbar}\Big)\Big).
	\label{eqS8}
\end{align}
And for $\rho_{\mathbf{n}_L,\mathbf{m}_L}$ both time-dependent perturbation contributions above should be considered.

\begin{figure}
	\centering
	\includegraphics[width=\textwidth]{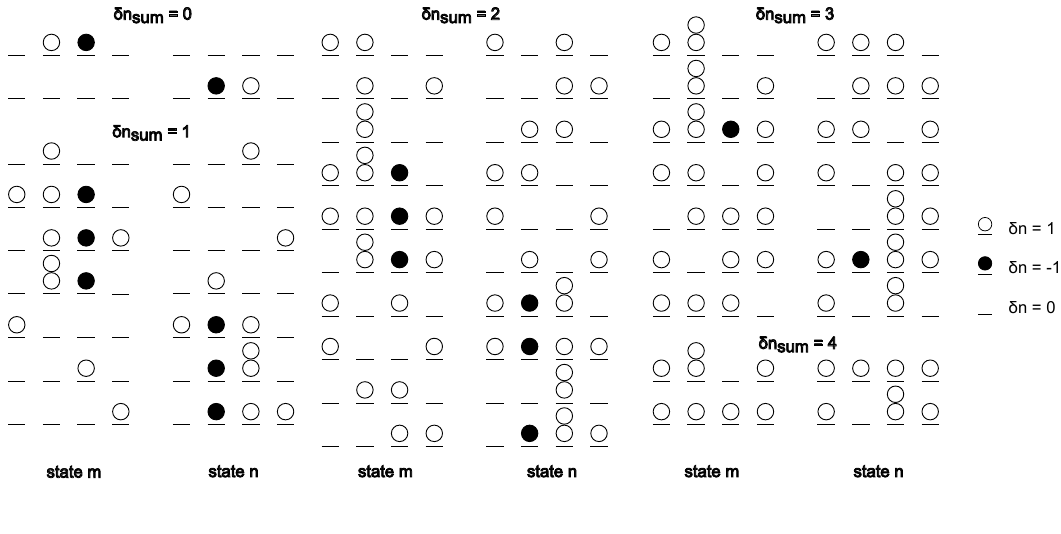}
	\caption{Possible correlation pairs of $\langle\mathbf{m}|\hat a^{\dagger}_2\hat a_3 |\mathbf{n}\rangle\neq0$, where states are all Fock states with a certain occupation number background $n_0$.}
	\label{figS8}
\end{figure}

Thus, all the correlation pairs in which we are interested contain at least one low-energy state, corresponding to $\rho_{\mathbf{n}_L,\mathbf{m}_L}(t)$, $\rho_{\mathbf{n}_L,\mathbf{m}_H}(t)$ and $\rho_{\mathbf{n}_H,\mathbf{m}_L}(t)$. We list all the possible correlation pairs of $\langle\mathbf{m}|\hat a^{\dagger}_2\hat a_3 |\mathbf{n}\rangle\neq0$ in Fig.~\ref{figS8}. So the next step is to calculate all these contributions and add up all the terms with the same frequencies. It seems like inevitable to walk through all the involved terms. However, mirror $\mathbb{Z}_2$ symmetry can do us a favor and simplify the form a lot. We can articulate this by classifying the categories of $|\mathbf{m}\rangle$ and $|\mathbf{n}\rangle$ into two cases: Case 1 is that one of them is high-energy state and the other is low-enegy state, Case 2 is that both of them are low-energy state

\vspace{0.3cm}

\textbf{Case 1. High-energy state and low-energy state}

\vspace{0.2cm}

Perturbative terms contributing to $\rho_{\mathbf{n}_L,\mathbf{m}_H}(t)$ are Eqs.~\eqref{eqS5} and \eqref{eqS7}. The dominant terms are $E^{i(0)}_{\mathbf{m}_H} - E^{i(0)}_{\mathbf{q}_L}=U_i$ and $E^{f(0)}_{\mathbf{m}_H} - E^{f(0)}_{\mathbf{q}_L}=U_f\pm E$, where the sign of $E$ is decided by whether $|\mathbf{q}_L\rangle$ and $|\mathbf{m}_H\rangle$ are coupled by $\hat{a}^{\dagger}_j\hat{a}_{j+1}$ or the hermitian conjugate. 
So we can simplify Eq.~\eqref{eqS5} as $\alpha_k (J_i/U_i)\mathrm{exp}(i(U_f-E)t/\hbar)$ and Eq.~\eqref{eqS7} as $\alpha_k(J_f/(U_f\pm E))(\mathrm{exp}(i(\mp E-E)t/\hbar)-\mathrm{exp}(i(U_f-E)t/\hbar))$, where $\alpha_k=\langle\mathbf{q}_L|\sum_{j=1}^{3}\hat{a}^{\dagger}_j\hat{a}_{j+1}+h.c.|\mathbf{m}_H\rangle\langle\mathbf{m}_H|\hat a^{\dagger}_2\hat a_3 |\mathbf{n}_L\rangle\rho^{(0)}_{\mathbf{n}_L,\mathbf{q}_L}$ and $k$ refers to the index set $\{\mathbf{m}_H,\mathbf{q}_L,\mathbf{n}_L\}$

Here is the key point: there exists a unique mirrored configuration $\rho_{\mathbf{\bar{m}}_H,\mathbf{\bar{n}}_L}(t)$ which has a conjugate coefficient $\alpha^{*}$. The corresponding perturbative terms are Eqs.~\eqref{eqS6} and \eqref{eqS8}, which can be simplified as $\alpha^{*}_k(J_i/U_i)\mathrm{exp}(i(-U_f-E)t/\hbar)$ and $\alpha^{*}_k(J_f/(U_f\mp E))(\mathrm{exp}(i(\mp E-E)t/\hbar)-\mathrm{exp}(i(-U_f-E)t/\hbar))$

\vspace{0.3cm}

\textbf{Case 2. Low-energy state and low-energy state}

\vspace{0.2cm}

Zero-order term of $\rho_{\mathbf{n}_L,\mathbf{m}_L}(t)$ is $\rho^{(0)}_{\mathbf{n}_L,\mathbf{m}_L}\mathrm{exp}(\frac{-iEt}{\hbar})$. Perturbative terms contributing to $\rho_{\mathbf{n}_L,\mathbf{m}_L}(t)$ are Eqs.~\eqref{eqS7} and \eqref{eqS8}. $E^{f(0)}_{\mathbf{m}_L} - E^{f(0)}_{\mathbf{q}_L}=\pm E$, where the sign of $E$ is decided by whether $|\mathbf{n}_L\rangle$ and $|\mathbf{m}_L\rangle$ are coupled by $\hat{a}^{\dagger}_j\hat{a}_{j+1}$ or the hermitian conjugate. So we can simplify Eq.~\eqref{eqS7} as $\beta_k(J_f/(\pm E))(\mathrm{exp}(i(\mp E-E)t/\hbar)-\mathrm{exp}(-iEt/\hbar))$
and Eqs.~\eqref{eqS8} as $\gamma_k(J_f/(\pm E))(\mathrm{exp}(i(\pm E-E)t/\hbar)-\mathrm{exp}({-iEt/\hbar}))$, where $\beta_k=\langle\mathbf{q}_L|\sum_{j=1}^{3}\hat{a}^{\dagger}_j\hat{a}_{j+1}+h.c.|\mathbf{m}_L\rangle\langle\mathbf{m}_L|\hat a^{\dagger}_2\hat a_3 |\mathbf{n}_L\rangle\rho^{(0)}_{\mathbf{n}_L,\mathbf{q}_L}$, $\gamma_k=\langle\mathbf{n}_L|\sum_{j=1}^{3}\hat{a}^{\dagger}_j\hat{a}_{j+1}+h.c.|\mathbf{q}_L\rangle \langle\mathbf{m}_L|\hat a^{\dagger}_2\hat a_3 |\mathbf{n}_L\rangle \rho^{(0)}_{\mathbf{q}_L,\mathbf{m}_L}$ and $k$ refers to the index set $\{\mathbf{m}_L,\mathbf{q}_L,\mathbf{n}_L\}$

The unique mirrored configuration $\rho_{\mathbf{\bar{m}}_L,\mathbf{\bar{n}}_L}$ which has a conjugate coefficient $\beta^{*}_k$ and $\gamma^{*}_k$. The corresponding perturbative terms are Eqs.~\eqref{eqS7} and \eqref{eqS8}, which can be simplified as $\gamma_k^{*}(J_f/(\mp E))(\mathrm{exp}(i(\pm E-E)t/\hbar)-\mathrm{exp}(-iEt/\hbar))$, and $\beta_k^{*}(J_f/(\mp E))(\mathrm{exp}(i(\mp E-E)t/\hbar)-\mathrm{exp}(-iEt/\hbar))$

Because all the matrix elements of $H^{(0)}$ are real, we can choose the bases such that $\rho^{(0)}_{\mathbf{n}_L,\mathbf{m}_L}\in \mathbb{R}$. Thus $\alpha_k,\beta_k,\gamma_k \in \mathbb{R}$, which results in all the perturbative terms in the "low-energy state and low-energy state" case cancel out. By summing all these terms, we get $G^{(1)}(1)=B\exp(iU_ft/\hbar)+A_{U+}\exp(iU_ft/\hbar)+A_{U-}\exp(-iU_ft/\hbar)+A_{E+}\exp(iEt/\hbar)+A_{E-}\exp(-iEt/\hbar)$, where
\[A_{U+}=\sum_{E^{f(0)}_{\mathbf{m}_H} - E^{f(0)}_{\mathbf{q}_L}=U_f+ E}\alpha_k\Big(\frac{J_i}{U_i}-\frac{J_f}{U_f + E}\Big)+\sum_{E^{f(0)}_{\mathbf{m}_H} - E^{f(0)}_{\mathbf{q}_L}=U_f- E}\alpha_k\Big(\frac{J_i}{U_i}-\frac{J_f}{U_f - E}\Big)\]
\[A_{U-}=\sum_{E^{f(0)}_{\mathbf{m}_H} - E^{f(0)}_{\mathbf{q}_L}=U_f+ E}\alpha_k\Big(\frac{J_i}{U_i}-\frac{J_f}{U_f - E}\Big)+\sum_{E^{f(0)}_{\mathbf{m}_H} - E^{f(0)}_{\mathbf{q}_L}=U_f- E}\alpha_k\Big(\frac{J_i}{U_i}-\frac{J_f}{U_f + E}\Big)\]
\[A_{E+}=\sum_{E^{f(0)}_{\mathbf{m}_H} - E^{f(0)}_{\mathbf{q}_L}=U_f-E}\alpha_k\frac{J_f}{U_f - E}+\alpha_k\frac{J_f}{U_f + E}
\]
\[A_{E-}=\sum_{E^{f(0)}_{\mathbf{m}_H} - E^{f(0)}_{\mathbf{q}_L}=U_f+E}\alpha_k\frac{J_f}{U_f+E}+\alpha_k\frac{J_f}{U_f - E}\]
\[   B=\sum_{\mathbf{m}_L,\mathbf{n}_L}\rho^{(0)}_{\mathbf{m}_L,\mathbf{n}_L}
\]

And $A_{U+},A_{U-},A_{E+},A_{E-}$ are small parameters compared to $B$, So $|G^{(1)}(1)|\approx B\,+(A_{U+}+A_{U-})\cos(U_ft/\hbar)\,+(A_{E+}+A_{E-})\cos(Et/\hbar)=B + A_E \cos\left(Et/\hbar\right) + A_{U} \cos\left(U_ft/\hbar\right)$, where
\[
A_U = C_{U}\frac{J_i}{U_i} - \frac{C_{U}}{2}\frac{J_f}{U_f-E} - \frac{C_{U}}{2}\frac{J_f}{U_f+E},
\]
\[
A_E = \frac{C_{U}}{2}\frac{J_f}{U_f-E} + \frac{C_{U}}{2}\frac{J_f}{U_f+E},
\]
\[
C_U= 2\sum_k\alpha_k.
\]

\subsection*{Dependence of the $E$-mode amplitude $A_E$ on $J_i/U_i$}
From the derivations above, we obtain quantitative expressions for both $A_U$ and $A_E$. To first order in perturbation theory, $A_E$ is independent of $J_i/U_i$. However, based on both numerical simulations and experimental data, we observe that $A_E$ exhibits a weak dependence on $J_i/U_i$, much weaker than that of $A_U$. In the inset of Fig.~4, we plot the rescaled $A_U$ and $A_E$ obtained from numerical simulations of a system without harmonic confinement. By linearly fitting $A_{E(U)}$ as functions of $J_i/U_i$, we extract the coefficients $C_E=0.100(5)$ and $C_U=0.760(11)$, respectively. The finite value of $C_E$ suggests a contribution beyond first-order perturbation, indicating a second-order effect.

\begin{figure}
	\centering
	\includegraphics[width=\textwidth]{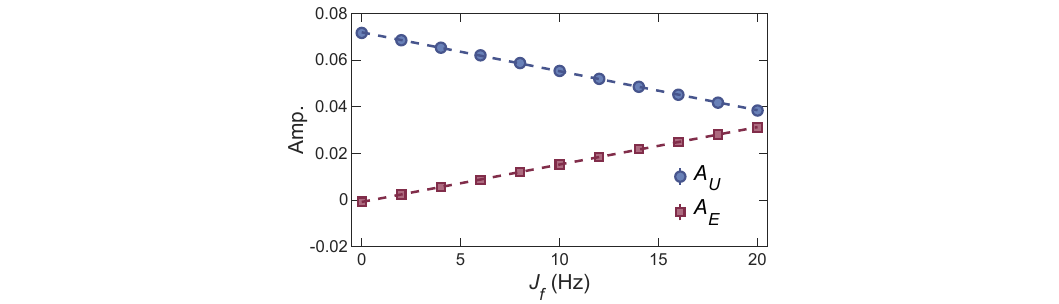}
	\caption{\textbf{Dependence of two-mode amplitudes $A_E$ and $A_U$ on $J_f$.} 
		Numerical simulation results for $A_E$ (red squares) and $A_U$ (blue circles) as functions of $J_f$. Both amplitudes are extracted from $\fc$ by fitting to Eq.~(2) (see \ref{S3}). Error bars are smaller than the marker size. The parameters are $J_i=80\ \mathrm{Hz}$, $U_i=909\ \mathrm{Hz}$, $U_f=1105\ \mathrm{Hz}$, $E=772\ \mathrm{Hz}$, and $\omega=0$.}
	\label{figS9}
\end{figure}

We now analyze the form of this second-order contribution. As discussed before, $A_E$ vanishes when $J_f=0$. Phenomenologically, this implies that the second-order term should scale as $J_f J_i$, consistent with a second-order process. To verify this, we examine the dependence of $A_{E(U)}$ on $J_f$ using numerical simulations without harmonic confinement. The system is quenched from $J_i=80\ \mathrm{Hz}$ and $U_i=909\ \mathrm{Hz}$ to $U_f=1105\ \mathrm{Hz}$ at $E=772\ \mathrm{Hz}$, with $J_f$ varied from $0$ to $20\ \mathrm{Hz}$. As shown in Fig.~\ref{figS9}, the amplitudes $A_E$ (red squares) and $A_U$ (blue circles) exhibit linear dependence on $J_f$, with $A_E \propto J_f$. This behavior is captured by the first-order contribution $\pm C_U[J_f/(U_f-E)+J_f/(U_f+E)]/2$ in $A_{E(U)}$, with the plus (minus) sign corresponding to $A_E$ ($A_U$). Linear fits yield slopes of $A_U$ and $A_E$ with respect to $J_f$ as $s_U=-s_E=-1.6\times10^{-3}$, in agreement with the model described in \ref{S4}.

Taking into account the constraint $A_E\propto J_f$, we conclude that the second-order term in $A_E$ takes the form $C'_E J_f J_i/(\Er U_i)$. Although this term may also depend on additional parameters such as $U_f$ and $E$, such contributions are weak and typically masked by the dominant first-order terms. As a result, the linear scaling of $A_E-(A_E)_{\rm c}$ with respect to $J_i/U_i-(J_i/U_i)_{\rm c}$ is only minimally affected.

\end{document}